\documentclass[onecolumn, superscriptaddress, aps, pra, 10pt, notitlepage, nofootinbib]{revtex4-2}

\usepackage{amsmath}
\usepackage{amsthm}
\usepackage{amssymb}
\usepackage[english]{babel}
\usepackage{dsfont}
\usepackage{enumitem}
\usepackage[margin=2cm]{geometry}
\usepackage{graphicx}
\usepackage[colorlinks=True, citecolor=blue, urlcolor=blue]{hyperref}
\usepackage{mathtools}
\usepackage{natbib}
\usepackage{wisjab}
\usepackage{xcolor}

\usepackage{lipsum}

\newtheorem{theorem}{Theorem}
\newtheorem{lemma}{Lemma}

\newtheorem{proposition}{Proposition}
\theoremstyle{definition}

\newenvironment{proofsketch}{\begin{proof}[Proof sketch.]}{\end{proof}}

\DeclareMathOperator{\tr}{tr}
\DeclareMathOperator{\sgn}{sgn}

\DeclareMathOperator{\polylog}{polylog}
\DeclareMathOperator{\argmax}{arg\>max}

\def\d{\mathrm d}
\def\ee{\mathrm e}
\def\ii{\kern0.05em\mathrm i\kern0.05em}

\def\rhom{\mtg\rho}

\def\ti{{t_{\rm i}}}
\def\tf{{t_{\rm f}}}
\def\lambdai{{\lambda_{\rm i}}}
\def\lambdaf{{\lambda_{\rm f}}}

\def\A{\mt A}
\def\H{\mt H}
\def\U{\mt U}
\def\dH{\partial_\lambda \H}
\def\cU{\mathcal U}
\def\cUt{\tilde{\mathcal U}}
\def\Hi{\H_{\rm i}}
\def\Hp{\H_{\rm p}}

\def\>{\kern0.08em}
\def\<{\kern-0.08em}

\begin{document}

\title{Gate-based counterdiabatic driving with complexity guarantees}

\author{Dyon van Vreumingen}
\affiliation{Institute of Physics, University of Amsterdam, Science Park 904, 1098 XH Amsterdam, The Netherlands}%
\affiliation{QuSoft, Centrum Wiskunde \& Informatica (CWI), Science Park 123, 1098 XG Amsterdam, The Netherlands}%

\date{\today}

\begin{abstract}
    We propose a general, fully gate-based quantum algorithm for counterdiabatic driving. The algorithm does not depend on heuristics as in previous variational methods, and exploits regularisation of the adiabatic gauge potential to suppress only the transitions from the eigenstate of interest. This allows for a rigorous quantum gate complexity upper bound in terms of the minimum gap $\Delta$ around this eigenstate. We find that, in the worst case, the algorithm requires at most $\tilde O(\Delta^{-(3 + o(1))} \epsilon^{-(1 + o(1))})$ quantum gates to achieve a target state fidelity of at least $1 - \epsilon^2$, where $\Delta$ is the minimum spectral gap. In certain cases, the gap dependence can be improved to quadratic.
\end{abstract}

\maketitle

\section{Introduction}
Adiabatic quantum computing (AQC) is a paradigm of quantum computation that leverages the principles of adiabatic processes in quantum mechanics to solve computational problems \cite{Albash2018}. AQC relies on the adiabatic theorem, which states that a quantum system approximately remains in its instantaneous eigenstate if a given time-dependent hamiltonian that governs its energy levels is changed slowly enough \cite{Kato1950}. In this sense, AQC is a conceptually straightforward way to prepare ground states or other eigenstates of complex systems, by starting from an eigenstate of a simple hamiltonian (e.g. a product state) and slowly time evolving towards the complex hamiltonian. \par
Contrary to AQC, counterdiabatic driving (CD) aims to prepare target states through a fast time evolution, actively suppressing excitations that would typically accompany such rapid driving. Traditional adiabatic processes require slow evolution to maintain the system in its instantaneous eigenstate, which can be impractically long. CD involves adding auxiliary, non-adiabatic control fields to counteract the diabatic transitions, thereby simulating the effect of a slow adiabatic process in a (much) shorter time \cite{Kolodrubetz2017}. It is a specific method within a broader class of techniques known as ``shortcuts to adiabaticity'' \cite{GueryOdelin2019}, designed to achieve the same goal of mimicking adiabatic evolution on a small time scale. Its original formulation is due to Demirplak \& Rice \cite{Demirplak2003, Demirplak2005, Demirplak2008}; in this formalism, the central object is the adiabatic gauge potential (AGP), which is an auxiliary field that is added to the system hamiltonian and cancels what is akin to a Coriolis force in the rotating system eigenbasis, thereby supressing all excitations. The issue however is that determining an AGP exactly for a given system is computationally difficult -- possibly as difficult, if not more, as preparing the eigenstate of interest -- and analytical expressions are only known for very specific systems \cite{Kolodrubetz2017}. This is due to the fact that, typically, the AGP is highly nonlocal and contains high-rank (two-body, three-body etc.) interactions \cite{delCampo2012, Saberi2014, Takahashi2013, Damski2014}; moreover, if the spectrum of the system is gapless, the AGP does not exist \cite{Jarzynski1995}. \par
For this reason, much effort has been focussed on computational methods that approximate the AGP. In this context, a pivotal role has been played by variational methods \cite{Sels2017}; these methods assume a parametrised ansatz that is optimised with respect to an action function that approaches zero as the ansatz approaches the exact AGP. This approach has enabled approximate CD in a large number of different contexts, leading to improved fidelities as compared to standard AQC \cite{Zhou2020, Passarelli2020, Xie2022, Hartmann2022, Mbeng2022, Cepaite2023, Schindler2023}. Among these variational methods, gate-based (digitised) approaches have been proposed as well \cite{Hegade2021}, where the dynamics is trotterised for implementation on a gate-based quantum computer. However, the limit to the variational approach lies in the fact that evaluating the action function itself becomes harder when higher-rank (two-body, three-body etc.) interactions are taken into account in the ansatz; therefore only ans\"atze with low-rank interactions are computationally feasible for variational methods. \par
Further improvements to this scheme were made through the invention of a nested commutator ansatz to the AGP \cite{Claeys2019} and subsequent Krylov space methods to solve for the optimal coefficients in this ansatz \cite{Bhattacharjee2023, Takahashi2024}.
Here, Krylov spaces are subspaces of operator space that are specifically chosen such that the coefficients can be found by solving a linear equation, which avoids optimisation heuristics with unpredictable running times. At the same time, this approach sheds more light on the computational complexity of computing the AGP. Numerical experiments \cite{Bhattacharjee2023} suggest that the relevant search space for weakly interacting and integrable systems is a small subspace in operator space, so that the AGP is easily approximated with an ansatz containing few coefficients; on the other hand, for strongly interacting, chaotic systems, this tends not to be the case, and the whole operator space may need to be explored. In this sense, the complexity of CD is tightly connected to quantum Krylov complexity \cite{Parker2019, Barbon2019, Bhattacharjee2022, Rabinovici2021, Rabinovici2022, Nandy2024}. More recently, interesting suggestions have been put forward to apply tensor network approaches for approximating the gauge potential by expressing it as a matrix product operator \cite{Kim2024, Keever2024}. \par
The main issue, however, in these developments, is that a precise description of the computational complexity of CD has so far been lacking. In particular, while the implementation of the aforementioned variational methods is straightforward, none of these have been accompanied with any rigorous computational complexity statement. The reason for this is twofold: (i) these approaches do not present any clear-cut quantification of the accuracy to which an AGP for a given system must be approximated in order to achieve a desired fidelity; (ii) the use of optimisation schemes makes complexity analysis of several of the methods prohibitively difficult. \par
In this work, address the complexity issue by asking the following question: does the computational effort we put into determining the AGP, in order to boost the fidelity of the output state with the target state, compare favourably to that of AQC with simply a longer evolution time? Following up on this question, how should we define computational complexity so that a fair comparison is made between the two methods? After all, the complexity of AQC is typically measured in terms of physical evolution time, whereas that of CD, in the light of aforementioned variational and Krylov space methods, boils down to the computing time a classical machine requires to solve a (linear) equation for the ansatz coefficients. To address this question, we place AQC and CD on an equal footing by considering the complexity of both approaches in terms of the number of quantum gates as required by a fault-tolerant quantum computer. To this end, we develop a gate-based quantum algorithm for CD (sections \ref{sec:reg_trunc_AGP} and \ref{sec:gate_based_CD}) that can be compared to gate-based version AQC (section \ref{sec:gate_based_AQC}). This method is nonvariational, and differs from the variational digitised approach previously put forward \cite{Hegade2021} in that it evades the difficulty of evaluating complexity that comes with variational methods. We give upper bounds on the number of quantum gates required for both gate-based CD and gate-based AQC. In this way, we contribute to the answering of the posed questions. \par
A crucial insight here is that the complexity of AQC is state dependent, in the sense that it is known to scale with the inverse of the spectral gap around the instantaneous eigenstate of interest, while that of currently known approaches to CD is not. So far, CD approaches have aimed to approximate the exact AGP, which cancels excitations on the entire spectrum; however, if the objective is to prepare only a single eigenstate, this is inefficient, since only the excitations from the eigenstate of interest need to be suppressed. We remedy this by basing our CD algorithm on an approximate AGP that effectively ignores excitations below a certain energy gap cutoff; this cutoff is then chosen small enough so that at least the transitions from the relevant eigenstate are suppressed. In this way, the gap around this eigenstate enters into the complexity description of CD. While the idea of an energy cutoff to suppress large-gap transitions itself is not new, and appears for example in the theory of quasi-adiabatic continuation \cite{Hastings2005, Hastings2010, Marien2014}, it has so far been overlooked in studies on the computational complexity of implementing counterdiabatic driving. \par
The main contribution of this paper is theorem \ref{thm:gate_based_CD}, which states the gate-based CD algorithm and an upper bound on its worst-case quantum gate complexity. The gate complexity bound we find is $\tilde O(\Delta^{-(3 + o(1))}\epsilon^{-(1 + o(1))})$, where $\Delta$ is the minimum energy gap along the path, and $\epsilon$ is the square root of the infidelity between the output state and the target state\footnote{The notation $f \in o(g)$ means that $\lim_{x\to\infty} f(x) / g(x) = 0$. Specifically, $o(1)$ is used to indicate a quantity that can be made arbitrarily small by increasing $x$. Additionally, we write $f \in \tilde O(g)$ to indicate $f \in O(g \polylog(g))$, $f \in \Omega(g)$ if $\lim_{x\to\infty} f(x) / g(x) > 0$ and $\Theta(g) = O(g) \cap \Omega(g)$.}.
Since $\Delta$ often decreases exponentially in the system size, especially in complex systems, the $\Delta$ dependence is usually the dominant factor in the complexity of AQC. The formalism we introduce in the work allows for the evaluation of CD in terms of $\Delta$ as well. \par
This work is structured as follows. In section \ref{sec:preliminaries}, we discuss the formalism of counterdiabatic driving and set out notational conventions used throughout this work. We also detail the Lie-Suzuki-Trotter (LTS) decomposition of a path-ordered exponential that translates a physical time description to a gate-based description. Section \ref{sec:reg_trunc_AGP} introduces the regularised truncated AGP, which brings the aforementioned gap cutoff into play and is at the heart of the gate-based CD algorithm. In section \ref{sec:gate_based_CD}, we construct the gate-based CD algorithm, which is a simulation, in the form of a LTS decomposition, of the path-ordered exponentiation of the regularised truncated AGP. In section \ref{sec:gate_based_AQC}, we compare the CD result to gate-based AQC by applying LTS simulation to standard adiabatic theorems. We conclude in section \ref{sec:conclusion_outlook}. For lemmas \ref{lem:regularised_truncated_unitary_error}--\ref{lem:LTS_error} and theorem \ref{thm:gate_based_CD}, we provide concise proof sketches in the main text, with the details deferred to appendices \ref{app:proof_reg_trunc_unit_err}--\ref{app:proof_gate_based_CD}. Besides the main body of work, appendix \ref{app:qdrift} discusses a randomised approach to gate-based CD which, depite its inferior complexity upper bound, may be of independent interest.

\section{Preliminaries}\label{sec:preliminaries}
\subsection{Counterdiabatic driving}
Consider the time-dependent Schr\"odinger equation for a system parametrised by a time-dependent parameter $\lambda(t)$:
\begin{align}
    \ii \> \d_t \ket{\psi(t)} = \H\big(\lambda(t)\big) \ket{\psi(t)}.
\end{align}
Clearly, the solution $\ket{\psi(t)}$ will be dependent on $\lambda(t)$. Let $\mt U\big(\lambda(t), \lambda(0)\big)$ be the unitary that connects the instantaneous eigenstates $\ket{n\big(\lambda(t)\big)}$ of $\H\big(\lambda(t)\big)$ with those of $\H\big(\lambda(0)\big)$ -- that is, let $\H\big(\lambda(t)\big)$ admit the spectral decomposition
\begin{align}
    \H\big(\lambda(t)\big) &= \sum_n E_n\big(\lambda(t)\big) \, \ketbra{n\big(\lambda(t)\big)} \nonumber \\
    &= \sum_n E_n\big(\lambda(t)\big) \, \mt U\big(\lambda(t), \lambda(0)\big) \ketbra{n\big(\lambda(0)\big)} \mt U\t\big(\lambda(t), \lambda(0)\big).
\end{align}
with $E_n\big(\lambda(t)\big)$ the corresponding energies (eigenvalues). We can view this system in a rotating ($\lambda$-dependent) basis by writing $\ket{\psi(t)} = \mt U\big(\lambda(t), \lambda(0)\big)\ket{\tilde\psi}$, where $\ket{\tilde\psi}$ is some reference state that is independent of $\lambda$ and $t$. In the rotating basis then, the time-dependent Schr\"odinger equation becomes
\begin{align}
    \ii \> \d_t \ket{\tilde\psi} = (\mt U\t \H \mt U - \ii \dot\lambda \mt U\t \partial_\lambda \mt U) \ket{\tilde\psi}. \label{eq:schr_eq_eff_hamil_rot_frame}
\end{align}
Since $\mt U\t \H \mt U$ is diagonal in the rotating eigenbasis of $\H$, any nonadiabatic transitions can only be generated by the second term of the effective hamiltonian in eq. \ref{eq:schr_eq_eff_hamil_rot_frame}. Thus we can suppress all nonadiabatic transitions by time evolving with a different hamiltonian that cancels out this second term in the rotating frame:
\begin{align}
    \mt U\t \H_{\rm CD} \mt U = \mt U\t \H \mt U + \ii \dot\lambda \mt U\t \partial_\lambda \mt U,
\end{align}
or, going back to the fixed ($\lambda$-independent) basis -- i.e. conjugating with $\mt U$ on both sides, we find
\begin{align}
    \H_{\rm CD}(t) = \H\big(\lambda(t)\big) + \dot\lambda(t) \A\big(\lambda(t)
    \big)
\end{align}
where the operator $\A = \ii \partial_\lambda \mt U \mt U\t$ is called the \emph{adiabatic gauge potential} (AGP). Since evolution under $\H_{\rm CD}$, known as the \emph{counterdiabatic hamiltonian}, causes no transitions into excited states, one may carry out a perfect evolution at any speed:
\begin{align}\label{eq:evolution_H_CD}
    \mt U(\lambdaf, \lambdai) = \mathcal T \exp \bigg[ \! -\!\ii \int_0^T \d t \, \H_{\rm CD}(t) \bigg] \quad\quad\quad \text{for any $T \in [0, \infty)$}.
\end{align}
The adiabatic gauge potential, then, can be viewed simply as the derivative operator $\ii \partial_\lambda$:
\begin{align}
    \bra{m(\lambda)} \mt A(\lambda) \ket{n(\lambda)} &= \ii \bra{m(\lambda)} \partial_\lambda \mt U(\lambda, \lambdai) \ket{n(\lambdai)} \nonumber \\
    &= \ii \bra{m(\lambda)} \partial_\lambda \ket{n(\lambda)},
\end{align}
and therefore generates motion in $\lambda$ space,
\begin{align}\label{eq:generated_parameter_translation}
    \mt U(\lambdaf, \lambdai) = \mathcal T \exp \bigg[ \! -\!\ii \int_{\lambdai}^{\lambdaf} \d\lambda \, \mt A(\lambda) \bigg]
\end{align}
which is simply a special case of eq. \ref{eq:evolution_H_CD}, namely the limit $T\to0$. From the Hellmann-Feynman theorem, $\mt A$ may be expressed in the eigenbasis of $\H(\lambda)$, and has the off-diagonal matrix elements
\begin{align}\label{eq:AGP_Hellmann_Feynman}
\bra{m(\lambda)}\mt A(\lambda)\ket{n(\lambda)} = \frac{\bra{m(\lambda)} \partial_\lambda \H(\lambda) \ket{n(\lambda)}}{\ii \omega_{mn}(\lambda)}
\end{align}
where $\omega_{mn} = E_m(\lambda) - E_n(\lambda)$. The diagonal elements are arbitrary since they depend on a gauge choice (namely, the phase of the eigenstates of $\H(\lambda)$) -- this makes the AGP nonunique for a given hamiltonian.
Lastly, the form of the AGP in eq. \ref{eq:AGP_Hellmann_Feynman} may be expressed as a time integral \cite{Claeys2019, Pandey2020}
\begin{align}\label{eq:AGP_time_integral}
\mt A(\lambda) = \frac12 \lim_{\eta\to0} \int_{-\infty}^\infty \d\tau \, \ee^{-\eta|\tau|} \sgn(\tau) \, \ee^{-\ii \H(\lambda) \tau} \partial_\lambda \H(\lambda) \> \ee^{\ii \H(\lambda) \tau}
\end{align}
since
\begin{align} \label{eq:AGP_matrix_elems}
\bra{m(\lambda)} \mt A(\lambda) \ket{n(\lambda)} &= \frac12 \bra{m(\lambda)} \partial_\lambda \H(\lambda) \ket{n(\lambda)} \lim_{\eta\to0} \int_{-\infty}^\infty \d\tau \, \ee^{-\eta|\tau|} \sgn(\tau) \, \ee^{-\ii \> \omega_{mn}(\lambda) \> \tau} \nonumber \\
&= \bra{m(\lambda)} \partial_\lambda \H(\lambda) \ket{n(\lambda)} \lim_{\eta\to0} \frac{\omega_{mn}(\lambda)}{\ii(\omega_{mn}(\lambda)^2 + \eta^2)} \nonumber \\
&= \frac{\bra{m(\lambda)} \partial_\lambda \H(\lambda) \ket{n(\lambda)}}{\ii \omega_{mn}}.
\end{align}
Note that the limit $\eta\to0$ must be taken after integration for the integral to converge.

\subsection{Norms and notation}
In this work, we will make use of two matrix norms: the spectral norm $\|\cdot\|$ (also known as operator norm) and the trace norm $\|\cdot\|_1$. The spectral norm is defined as $\|\mt O\| = \sup_{\ket\psi: \|\psi\|=1} \|\mt O\ket\psi\|$, or equivalently as the largest singular value of $\mt O$. For hermitian matrices, this corresponds to the largest absolute eigenvalue. The trace norm is defined by $\|\mt O\|_1 = \tr\sqrt{\mt O\t\mt O}$, which equates to the sum over all singular values of $\mt O$ (or absolute eigenvalues if $\mt O$ is hermitian). Both norms satisfy a triangle inequality and are multiplicative, as well as unitarily invariant, in the sense that $\|\mt U \mt O \mt V\|_{(1)} = \|\mt O\|_{(1)}$ (using subscript $_{(1)}$ to indicate either spectral or trace norm) for unitary matrices $\mt U, \mt V$. The norms are related by the inequality $\|\mt O \mt Q\|_1 \leq \|\mt O\| \|\mt Q\|_1$ for any matrices $\mt O$ and $\mt Q$, which may be derived, for example, from Hölder's inequality or Von Neumann's inequality. Furthermore, both norms satisfy the so-called telescoping property: let $\mt U_1, \mt U_2, \mt V_1, \mt V_2$ be unitary matrices; then
\begin{align}
    \|\mt U_1 \mt U_2 - \mt V_1 \mt V_2\|_{(1)} \leq \|\mt U_1 - \mt V_1\|_{(1)} + \|\mt U_2 - \mt V_2\|_{(1)},
\end{align}
which follows from the unitary invariance of the norms and a triangle equality. From this it may be immediately seen that errors in a product of unitaries scale at most linearly in the number of factors. \par
The spectral and trace norms give rise to the spectral distance $\|\mt O - \mt Q\|$ and trace distance $\frac12 \|\mt O - \mt Q\|_1$ respectively. The trace distance is especially useful when dealing with density matrices. In particular, for pure states, it can be shown that
\begin{align}
    \frac12 \|\ketbra\psi - \ketbra\phi\|_1 = \sqrt{1 - |\braket\psi\phi|^2}.
\end{align}
The derivation follows from explicit calculation of the eigenvalues of $\ketbra\psi - \ketbra\phi$ after Gram-Schmidt orthogonalisation. We will call the right-hand side of this equation the square-root infidelity between $\ket\psi$ and $\ket\phi$. It may be shown that the square-root infidelity is upper bounded by the euclidean distance:
\begin{align}
    \sqrt{1 - |\braket\psi\phi|^2} \leq \|\ket\psi - \ket\phi\|.
\end{align}
Throughout this work, we will frequently encounter $\lambda$-dependent vector norm expressions of the form $\|\mt O(\lambda)\ket{n(\lambda)}\|$ where $\ket{n(\lambda)}$ is an instantaneous eigenstate of the system hamiltonian. To simplify notation, we will use the shorthand
\begin{align}
    \|\mt O\|_{n(\lambda)} := \|\mt O(\lambda)\ket{n(\lambda)}\|
\end{align}
for this purpose. When clear from context, the $\lambda$ argument will be omitted. Finally, we introduce convenient notation for the following integral expressions:
\begin{align}
    \|\mt O\|_{n, p} := \bigg( \int_\lambdai^\lambdaf \d\lambda \, \|\mt O(\lambda)\ket{n(\lambda}\|^p \bigg)^{1 / p} \;; \quad\quad\quad\quad \|\mt O\|_{\infty, p} := \bigg( \int_\lambdai^\lambdaf \d\lambda \, \|\mt O(\lambda)\|^p \bigg)^{1 / p}
\end{align}
for $p \in \mathbb N_{>0}$; $\|\mt O\|_{n, \infty}$ and $\|\mt O\|_{\infty, \infty}$ are understood to mean $\max_{\lambda\in[\lambdai, \lambdaf]} \|\mt O(\lambda)\ket{n(\lambda)}\|$ and $\max_{\lambda\in[\lambdai, \lambdaf]} \|\mt O(\lambda)\|$ respectively.

\subsection{Lie-Trotter-Suzuki decompositions}\label{sec:LTS}
Since the main objective of this work is to describe a counterdiabatic driving protocol that runs on a digital quantum computer, a significant portion of the paper will be dedicated to simulation -- that is, the approximation as a sequence of quantum gates -- of ordered exponentials. The problem of implementing ordered exponentials of hermitian operators is far from new: several time-dependent hamiltonian simulation routines exist for implementing ordered exponentials. The most notable of these are Lie-Trotter-Suzuki (LTS) decompositions \cite{Suzuki1993, Wiebe2010} and truncated Dyson series methods \cite{Low2019a, Kieferova2019}, though more approaches exist \cite{Berry2020, Watkins2022}. A typical requirement for these methods is that the hamiltonian be provided as a sum of finitely many terms $\H(t) = \sum_{i=1}^\ell \H_i(t)$; or more specifically, as a linear combination of unitaries (LCU), $\H(t) = \sum_{i = 1}^\ell \beta_i(t) \mt V_i$, where the $\mt V_i$ are unitary and the $\beta_i$ are nonnegative scalars\footnote{For many-body hamiltonians, these LCUs can be readily constructed as linear combinations of Pauli strings (tensor products of Pauli operators) through the Jordan-Wigner and Bravyi-Kitaev transforms, as well as more sophisticated methods \cite{Lee2021, Loaiza2024}.}. LTS formulae approximate a time-ordered exponential $\U(\tf, \ti) = \mathcal T \exp[-\ii \int_\ti^\tf \d t \H(t)]$ as a product $\U(\tf, \ti) \approx \prod_j \ee^{-\ii \H_{i_j}(t_j) \delta t_j}$. Each factor in this product may then be simulated through time-\emph{independent} hamiltonian simulation methods such as qubitisation \cite{Low2019b}, or in the case of LCU, as a multi-qubit rotation gate. Truncated Dyson series, on the other hand, seek to implement the approximation $\U(\tf, \ti) \approx \sum_{k = 0}^K \frac{(-\ii)^k}{k!} \int_\ti^\tf \cdots \int_\ti^\tf \mathcal T[\H(t_k) \cdots \H(t_1)] \d^k t$. In this process, oracle access to the hamiltonian is assumed, in the form of a time-independent and a time-dependent oracle, where the time-dependent oracle is a direct sum of time-independent oracles over a finite set of times in the simulation interval (see ref. \citenum{Low2019a} for details). While efficient methods have been developed to decompose the time-independent oracle into a sequence of quantum gates \cite{Shende2006, Childs2012, Childs2018, Low2019b}, such decompositions for the time-dependent oracle are only known in special cases \cite{Low2019a}. For this reason, we will use LTS formulae to describe gate-based AQC and CD throughout the rest of this work. \par
The starting point in the description of LTS formulae is the following observation: for time-independent operators $\mt O$ and $\mt Q$ and $\delta t \in \mathbb R$, the exponential $\ee^{-\ii (\mt O + \mt Q) \delta t}$ can be split into the product $\ee^{-\ii \mt O \delta t}\ee^{-\ii \mt Q \delta t}$ at the cost of an error $\|\ee^{-\ii (\mt O + \mt Q) \delta t} - \ee^{-\ii \mt O \delta t} \ee^{-\ii \mt Q \delta t}\| \in O(\delta t^2)$. Clearly, if the time step $\delta t$ is small, so is the error. If $\delta t$ is large, one may divide $\delta t$ into $r$ parts to obtain $\|\ee^{-\ii (\mt O + \mt Q) \delta t} - (\ee^{-\ii \mt O \delta t / r} \ee^{-\ii \mt Q \delta t / r})^r\| \in O(\delta t^2 / r)$. Here, the additional factor $r$ in the error follows from linear propagation of errors as a result of the telescoping property of the operator norm. The parameter $r$ can then be chosen to achieve any desired precision; namely, to upper bound the error by $\epsilon$, $r$ must be taken to be at least $O(\delta t^2 / \epsilon)$. This is the most elementary kind of Trotter formula, and forms the essence of all Trotter-based decompositions. \par
Lie-Trotter-Suzuki (LTS) formulae are more sophisticated decompositions, and are due to Suzuki \cite{Suzuki1993}. These decompositions may be defined for any order $k \in \mathbb N_{>0}$, with every order representing a more fine-grained approximation, and achieve $O(\delta t^{2k + 1})$ error scaling. Again dividing a large $\delta t$ into $r$ smaller segments, this results in an $r$ scaling $O(\delta t^{1 + 1 / 2k} \epsilon^{1 / 2k})$ to achieve precision $\epsilon$.\footnote{The Trotter decomposition $\ee^{-\ii (\mt O + \mt Q) \delta t} = (\ee^{-\ii \mt O \delta t / r} \ee^{-\ii \mt Q \delta t / r})^r + O(\delta t^2 / r)$ may be viewed as analogous to a LTS formula of order $k = 1/2$.} \par
So far, we have described the Trotter method only in relation to time-independent operator exponentials, but the concept naturally extends to time-dependent operator exponentials. While the generalisation of time-independent decompositions themselves to time-dependent ones is straightforward, general error bounds were found only years later by Wiebe et al. \cite{Wiebe2010}. Their result is very similar to the path-independent case, but imposes conditions on the differentiability of the operator exponents. We base the analysis in this section on their work, accurately describing the decomposition of the ordered exponential $\U(\tf, \ti) = \mathcal T \exp [-\ii \int_\ti^\tf \d t \, \H(t)]$ and the requirements to achieve a given precision. \par
The first-order single-segment LTS approximation $\tilde{\mt U}_{1, 1}(\tf, \ti)$ of $\U(\tf, \ti)$, for the operator $\H(t) = \sum_{i = 1}^\ell \H_i(t)$, is given by
\begin{align}\label{eq:tilde_U_1,1_definition}
    \tilde{\mt U}_{1, 1}(\tf, \ti) = \bigg( \prod_{i = 1}^\ell \exp [-\ii \> \H_i(\ti + \delta t_1 / 2) \> \delta t_1 / 2] \bigg) \bigg( \prod_{i = \ell}^1 \exp [-\ii \> \H_i(\ti + \delta t_1 / 2) \> \delta t_1 / 2] \bigg).
\end{align}
where $\delta t_1 = \tf - \ti$. Higher-order LTS product formulas are built by repeatedly applying the following recursive process to $\tilde{\mt U}_{1, 1}$:
\begin{align}\label{eq:LTS_order_k+1}
\tilde{\mt U}_{k + 1, 1}(t + \delta t, t) &= \tilde{\mt U}_{k, 1}(t + \delta t, t + [1 - s_k] \delta t) \, \tilde{\mt U}_{k, 1}(t + [1 - s_k] \delta t, t + [1 - 2 s_k] \delta t) \nonumber\\
&\quad\, \times \tilde{\mt U}_{k, 1}(t + [1 - 2 s_k] \delta t, t + 2 s_k \delta t) \, \tilde{\mt U}_{k, 1}(t + 2 s_k \delta t, t + s_k \delta t) \, \tilde{\mt U}_{k, 1}(t + s_k \delta t, t)
\end{align}
with $s_k = (4 - 4^{1 /(2 k + 1)})^{-1}$. Now if every $\H_i(t)$ is $2k$ times differentiable in $t$, then the error is bounded as in the path-independent case:
\begin{align}
    \|\U(\tf, \ti) - \tilde\U_{k, 1}(\tf, \ti)\| \in O(\delta t^{2k + 1}).
\end{align}
This means that if $\delta t$ is small, then the approximation error decreases monotonically with the order $k$. If $\delta t$ is not small, we divide it into $r$ segments and apply a LTS product formula for each segment:
\begin{align}\label{eq:LTS_U_k_r}
    \tilde{\mt U}_{k, r}(t + \delta t, t) = \prod_{s = 1}^r \tilde{\mt U}_k(t + s \> \delta t / r, t + (s - 1) \delta t / r).
\end{align}
We refer to this decomposition as a $k$\textsuperscript{th}-order $r$-segment LTS formula. The work of Wiebe et al. \cite{Wiebe2010} now provides a condition on $r$ that is sufficient to upper bound the spectral distance between an ordered exponential of a sum of operators and its LTS simulation. 
\begin{lemma}[Wiebe et al. \cite{Wiebe2010}, theorem 1]\label{lem:Wiebe}
    Let $\mt H(t) = \sum_i \mt H_i(t)$ be defined on the interval $[\ti, \tf]$ such that each $\mt H_i(t)$ is hermitian and $2k$ times differentiable on the entire interval. Let $\delta t = \tf - \ti$. Furthermore, suppose
    \begin{align}\label{eq:Wiebe_Lambda}
        \max_{p = 0, 1, \ldots, 2k} \bigg[ \max_{t \in [\ti, \tf]} \bigg( \sum_i \bigg\| \frac{\partial^p}{\partial t^p} \mt H_i(t) \bigg\| \bigg)^{1 / (p + 1)} \bigg] \leq \Lambda.
    \end{align}
    If $\epsilon \leq \min\{(9/10)(5/3)^k \> \Lambda \> \delta t, 1\}$, then the spectral distance between the ordered exponential $\mathcal T\exp[-\ii \int_{\ti}^{\tf} \d t \, \mt H(t)]$ and its $k$\textsuperscript{th}-order $r$-segment LTS decomposition is at most $\epsilon$, provided that
    \begin{align}
        r \geq 5k \> \Lambda \> \delta t \> \bigg( \frac53 \bigg)^k \bigg( \frac{\Lambda \delta t}\epsilon \bigg)^{1/2k}.
    \end{align}
\end{lemma}

\section{Regularised truncated gauge potential}\label{sec:reg_trunc_AGP}
In the following sections, we will describe our approach to a gate-based counterdiabatic driving algorithm. We will base our methods on the time-integral expression for the adiabatic gauge potential from eq. \ref{eq:AGP_time_integral}. However, in its stead, we will use its regularised truncated version $\mt A_{\eta, a}$. Here, regularised means that we fix an $\eta > 0$ instead of taking the limit $\eta \to 0$; furthermore, we restrict the integration to a bounded interval $[-a, a]$. That is:
\begin{align}\label{eq:regularised_truncated_AGP}
    \mt A_{\eta, a}(\lambda) = \frac12 \int_{-a}^a \d\tau \, \ee^{-\eta|\tau|} \sgn(\tau) \> \ee^{-\ii \H(\lambda) \tau} \dH(\lambda) \> \ee^{\ii \H(\lambda) \tau}.
\end{align}
One may view this approximation choice as a straightforward way to get rid of unworkable infinities. But there is a more intuitive reason to regularise the AGP: considering the matrix elements of $\mt A$ in the energy eigenbasis (eq. \ref{eq:AGP_matrix_elems}), one would intuitively expect counterdiabatic driving with the regularised AGP to suppress only those level transitions where $\eta^2 \ll \omega_{mn}^2$ \cite{Pandey2020}. In this sense, $\eta$ can be viewed as a gap cutoff. Since we're only interested in suppressing the level transitions from the $n$-th eigenstate, a cutoff $\eta \sim \Delta_n^\nu$ for some $\nu > 0$, with $\Delta_n$ the minimum gap around the $n$-th eigenstate, should suffice.
This energy cutoff $\eta$ is also turns out to be related to the cost of implementing approximate CD with the regularised AGP: as we will show later, the algorithm complexities grow with $\eta^{-1}$ and $a$. Besides that, there is a connection (albeit qualitative) between $\eta^{-1}$ and the thermodynamic cost associated with CD \cite{Campbell2017, Zheng2016}. In these works, the norm of the AGP is introduced as a cost of implementing CD; at the same time, one can easily see that the spectral norm of the regularised truncated AGP satisfies $\|\mt A_{\eta, a}\| \leq \eta^{-1} (1 - \ee^{-\eta a}) \|\dH\| \leq \eta^{-1} \|\dH\|$. \par
For the remainder of the work, we will work in the limit $T\to0$ (eq. \ref{eq:generated_parameter_translation}). The most important reason for this is that, in the proofs of the error bounds that follow, the system hamiltonian always cancels out (see eq. \ref{eq:regularised_truncated_unitary_AGP_error} in the proof of lemma \ref{lem:regularised_truncated_unitary_error}). As such, working in the limit $T\to0$ and thereby leaving out the system hamiltonian merely simplifies the calculations. Nonetheless, one might argue that a better complexity can be achieved by choosing an optimal $T$ which is not reflected in the presented bounds. We expect that such an optimal $T^*$ would be at least on the order $T^* \in \Omega(\Delta^{-1})$, and possibly $T^* \in \Omega(\|\partial_\lambda \mt H\|_{\infty, \infty} / \Delta^2)$. (Larger times essentially revert the protocol to adiabatic evolution.) In such cases, we expect the added complexity of trotterising $\H$ to be asymptotically subdominant to that of CD in the limit $T \to 0$, since the complexity of AQC at a time scale $T$ grows almost linearly in $T$ (see proposition \ref{prop:gate_based_AQC} in section \ref{sec:gate_based_AQC}). However, we do not expect such schemes to provide significant gains over either our CD algorithm or AQC, since CD in the $T \to 0$ limit can be essentially be viewed as AQC in a different coordinate system (i.e. a coordinate change from time to $\lambda$) and a $T > 0$ setting in a sense interpolates between the two. \par
We proceed to establish a more precise relation between $\eta$, $a$ and $\Delta_n$, which puts conditions on $\eta$ and $a$ for arbitrary approximation error.

\begin{lemma}\label{lem:regularised_truncated_unitary_error}
Let $\U_{\eta, a}(\lambdaf, \lambdai) = \mathcal T \exp[-\ii \int_\lambdai^\lambdaf \d\lambda \> \A_{\eta, a}(\lambda)]$, where $\A_{\eta, a}(\lambda)$ is defined as in eq. \ref{eq:regularised_truncated_AGP}, and let $\ket{n(\lambdai)}$ be the $n$-th eigenstate of $\H(\lambdai)$. Then
\begin{align}
    \|(\U(\lambdaf, \lambdai) - \U_{\eta, a}(\lambdaf, \lambdai))\ket{n(\lambdai)}\| \leq \epsilon
\end{align}
if we take
\begin{align}\label{eq:eta_a_conditions}
\eta = \frac1{\sqrt2} \Delta_n^{3/2} \epsilon^{1/2} \|\dH\|_{n, 1}^{-1/2}; \quad\quad 
a = \frac1\eta \log\bigg( 2 \> \frac{\Delta_n + \eta}{\Delta_n\epsilon\eta} \|\dH\|_{n, 1} \bigg) \in \tilde O(\eta^{-1}).
\end{align}
\end{lemma}

\begin{proofsketch}
    We first express the error in terms of the exact and approximate AGP by bounding $\|(\U(\lambdaf, \lambdai) - \U_{\eta, a}(\lambdaf, \lambdai))\ket{n(\lambdai)}\| \leq \int_\lambdai^\lambdaf \d\lambda \> \|\A - \A_{\eta, a}\|_{n(\lambda)}$ with a triangle inequality. By expanding $\A(\lambda)$ and $\A_{\eta, a}(\lambda)$ in the instantaneous energy eigenbasis, we find that the absolute values of the off-diagonal matrix entries $\bra m(\A - \A_{\eta, a})\ket n$ are upper bounded by
    \begin{align}
        |\bra m(\A - \A_{\eta, a})\ket n| \leq |\bra m \dH \ket n| \bigg( \frac{\eta^2}{|E_m - E_n|^3} + O(\ee^{-\eta a}) \bigg).
    \end{align}
    The result then follows from a simple calculation.
\end{proofsketch}
The full proof is given in appendix \ref{app:proof_reg_trunc_unit_err}.

\section{Gate-based counterdiabatic driving}\label{sec:gate_based_CD}
With the regularised truncated AGP in hand, we can now discuss approaches to gate-based counterdiabatic driving. For this purpose, we will understand CD as the task of implementing the ordered exponential $\U_{\eta, a}(\lambdaf, \lambdai) = \mathcal T \exp[-\ii \int_\lambdai^\lambdaf \d\lambda \> \A_{\eta, a}(\lambda)]$ with some unitary $\tilde\U(\lambdaf, \lambdai)$, up to a given precision $\epsilon$; by a simple triangle equality, the total error $\|(\U(\lambdaf, \lambdai) - \tilde\U(\lambdaf, \lambdai))\ket{n(\lambdai)}\|$ is then at most $O(\epsilon)$. In this section, we present a deterministic gate-based algorithm that achieves exactly this goal, and provide an upper bound on the complexity of running this algorithm on a fault-tolerant gate-based quantum computer. \par
Evidently, time-dependent simulation of the ordered exponential $\U_{\eta, a}$ is an integral component of gate-based CD, and we will employ LTS formulae for this purpose, as described in section \ref{sec:LTS}. However, since the LTS formalism assumes that the integrand in the exponent is a sum of finitely many operator terms, and $\A_{\eta, a}$ is defined as an operator integral, we first need to approximate $\A_{\eta, a}$ with a suitable sum. To this end, we give a construction that is similar to an operator-valued Riemann approximation to the integral, except we make a more clever choice of evaluation points and weights way to obtain a smaller error bound. As such, all terms in the sum are proportional to the integrand in eq. \ref{eq:regularised_truncated_AGP}, i.e. an operator of the form $\exp[-\ii \H(\lambda) \tau] \> \dH(\lambda) \exp[\ii \H(\lambda) \tau]$. \par
This leads nother issue: the mentioned time-dependent hamiltonian simulation routines assume that the operator terms in the sum are given, in the same sense that we assume $\H(\lambda)$ and $\dH(\lambda)$ given as sums of simple (weighted unitary) terms. Clearly, this is not the case here, since one would first have to simulate all operators $\exp[\pm \ii \H(\lambda)\tau]$ to construct the desired operator terms. To avoid nested simulation, we should integrate the simulation of the $\exp[\pm \ii \H(\lambda)\tau]$ operators in the procedure for approximating $\mt U_{\eta, a}$. Fortunately, this can be done in a straightforward way using Trotter-based formulae, since when we exponentiate the integrand terms, the operators $\exp[\pm \ii \H(\lambda)\tau]$ may be taken out of the exponent:
\begin{align}
    \exp[ -\ii (\ee^{-\ii \H(\lambda) \tau} \dH(\lambda) \ee^{\ii \H(\lambda) \tau}) \delta\lambda] = \exp[-\ii \H(\lambda) \tau] \exp[-\ii \dH(\lambda) \delta\lambda] \exp[\ii \H(\lambda) \tau].
\end{align}
This follows immediately from the fact that $\exp[\U \mt O \U\t] = \U \exp[\mt O] \U\t$, for any $\mt O$ and unitary $\U$, which may be shown using a Taylor expansion. As such, we avoid double exponentiation and are left with only a product of ordinarily simulatable operator exponentials. \par
In what follows, we first give a detailed description of our summation method, which gives rise to an AGP approximation $\A_{\eta, a}^{M, q} \approx \A_{\eta, a}$ and a corresponding evolution operator $\U_{\eta, a}^{M, q}(\lambdaf, \lambdai) = \mathcal T\exp[-\ii \int_\lambdai^\lambdaf \d\lambda \> \A_{\eta, a}^{M, q}(\lambda)]$. Subsequently, we give a LTS construction to simulate $\U_{\eta, a}^{M, q}$, producing the unitary $\tilde\U$. We establish conditions on the relevant parameters to upper bound the errors $\|(\U_{\eta, a}(\lambdaf, \lambdai) - \U_{\eta, a}^{M, q}(\lambdaf, \lambdai))\ket{n(\lambdai)}\|$ and $\|(\U_{\eta, a}^{M, q}(\lambdaf, \lambdai) - \tilde\U(\lambdaf, \lambdai))\ket{n(\lambdai)}\|$; the error $\|(\U_{\eta, a}(\lambdaf, \lambdai) - \tilde\U(\lambdaf, \lambdai))\ket{n(\lambdai)}\|$ is then simply the sum of these two errors, by a triangle inequality. Finally, setting this error to at most $O(\epsilon)$ leads to a gate complexity expression for the algorithm.

\subsection{Weighted sum approximation}
The most elementary way of approximating an integral of a scalar-valued function is a Riemann sum: the integration range is partitioned into $M$ subintervals and one approximates $\int_a^b \d x \, f(x) \approx \sum_{\kappa = 1}^M f(x_\kappa) \delta x_\kappa$. A similar thing can be done with operator functions. Let us write
\begin{align}
    \mt A_{\eta, a} = \frac12 \int_0^a \d\tau \, \ee^{-\eta\tau} \big( \partial_\lambda \H(\tau) - \partial_\lambda \H(-\tau) \big)
\end{align}
where we dropped the $\lambda$ argument for legibility, and denote $\mt O(\tau) = \ee^{-\ii \H \tau} \mt O \> \ee^{\ii \H \tau}$ for any operator $\mt O$. We partition the interval $[0, a]$ into $M$ subintervals, picking $\tau_0 < \tau_1 < \cdots < \tau_M$ such that $\tau_0 = 0$ and $\tau_M = a$; define $\delta\tau_\kappa = \tau_\kappa - \tau_{\kappa - 1}$. Note that the subintervals need not be uniform. The operator Riemann sum then takes the form
\begin{align}
    \A_{\eta, a} \approx \frac12 \sum_{\kappa = 1}^M \delta\tau_\kappa \ee^{-\eta\tau_\kappa} \big( \dH(\tau_\kappa) - \dH(-\tau_\kappa) \big).
\end{align}
However, this is a rather crude method in that the error in each subinterval is relatively large, so that many subintervals are needed to make the overall error small. A more clever method evaluates the integrand at multiple points in each interval and weighs each term accordingly:
\begin{align}\label{eq:lagrange_interpolated_AGP}
    \mt A_{\eta, a} \approx \mt A_{\eta, a}^{M, q} = \frac12 \sum_{\kappa = 1}^M \delta\tau_\kappa \sum_{\alpha = 0}^q \> w_{\kappa, \alpha} \> \ee^{-\eta \tau_{\kappa, \alpha}} \big( \partial_{\lambda} \H(\tau_{\kappa, \alpha}) - \partial_\lambda \H(-\tau_{\kappa, \alpha}) \big).
\end{align}
Here, $q \leq 0$ is introduced as a free parameter, so that $q + 1$ evaluations are made within each subinterval. The weights $w_{\kappa, \alpha}$ are determined through Lagrange interpolation (see appendix \ref{app:lagrange_interpolation}) and are such that $\sum_{\alpha = 0}^q w_{\kappa, \alpha} = 1$ for all $\kappa$. Furthermore, they depend only on our choice of the interpolation points $\tau_{\kappa, \alpha}$ and not on the value of the integrand at those points. We will not specify the choice of the interpolation points, and instead rely on general Lagrange interpolation bounds for estimating the error made in the approximation. This gives rise to the following lemma.

\begin{lemma}\label{lem:summation_error}
Let $\U_{\eta, a}(\lambdaf, \lambdai)$ be given as in lemma \ref{lem:regularised_truncated_unitary_error}, let $\mt A_{\eta, a}^{M, q}(\lambda)$ be given as in eq. \ref{eq:lagrange_interpolated_AGP} and let $\U_{\eta, a}^{M, q}(\lambdaf, \lambdai) = \mathcal T\exp[-\ii \int_\lambdai^\lambdaf \d\lambda \> \A_{\eta, a}^{M, q}(\lambda)]$. Assume $\eta \leq \min_{\lambda\in[\lambdai, \lambdaf]} \|\H(\lambda)\|$. 
If we choose
\begin{align}\label{eq:tau_kappa}
    \tau_\kappa = -\frac{q + 2}\eta \log \< \Big( 1 - \frac\kappa M (1 - \ee^{-\eta a / (q + 2)}) \Big), \quad\quad \kappa \in \{0, \ldots, M\}
\end{align}
and if we set, for any choice of interpolation points $\tau_{\kappa, 0} < \tau_{\kappa, 1} < \cdots < \tau_{\kappa, q}$ within each subinterval $[\tau_{\kappa - 1}, \tau_{\kappa}]$,
\begin{align}\label{eq:w_kappa,alpha}
    w_{\kappa, \alpha} = \frac1{\delta\tau_\kappa} \int_{\tau_{\kappa - 1}}^{\tau_\kappa} \d\tau \prod_{\substack{0\leq\beta\leq q \\ \beta\neq\alpha}} \frac{\tau - \tau_{\kappa, \beta}}{\tau_{\kappa, \alpha} - \tau_{\kappa, \beta}}, \quad\quad \alpha\in\{0, \ldots, q\}
\end{align}
then
\begin{align}
    \|(\U_{\eta, a}(\lambdaf, \lambdai) - \U_{\eta, a}^{M, q}(\lambdaf, \lambdai))\ket{n(\lambdai)}\| \leq \epsilon
\end{align}
provided that
\begin{align}\label{eq:M_condition}
    M \geq \max\bigg\{ \frac{3 \> \ee \> (2a)^{1 + 1 / (q + 1)}}{\epsilon^{1 / (q + 1)} (q + 1)} \|\H\|_{\infty, \infty} \|\partial_\lambda \H\|_{n, 1}^{1 / (q + 1)}, \ee^{\eta a / (q + 2)} - 1 \bigg\}.
\end{align}
\end{lemma}

\begin{proofsketch}
    As before, we bound $\|(\U_{\eta, a}(\lambdaf, \lambdai) - \U_{\eta, a}^{M, q}(\lambdaf, \lambdai))\ket{n(\lambdai)}\| \leq \int_\lambdai^\lambdaf \d\lambda \> \|\A_{\eta, a} - \A_{\eta, a}^{M, q}\|_n$, and then apply another triangle inequality to bound this quantity by a sum of contributions from each of the $M$ subintervals:
    \begin{align}\label{eq:proofsketch_lem_3_step_1}
        \| \mt A_{\eta, a} - \mt A_{\eta, a}^{M, q} \|_n \leq \frac12 \sum_{\kappa=1}^M \bigg( \,
        \bigg\| \int_{\tau_{\kappa - 1}}^{\tau_\kappa} \d\tau \, \ee^{-\eta\tau} \> \partial_\lambda \H(\tau) - \sum_{\alpha = 0}^q \> w_{\kappa, \alpha} \> \ee^{-\eta \tau_{\kappa, \alpha}} \partial_{\lambda} \H(\tau_{\kappa, \alpha}) \bigg\|_n + (\tau \leftrightarrow -\tau) \bigg).
    \end{align}
    Given that the weights $w_{\kappa, \alpha}$ in eq. \ref{eq:w_kappa,alpha} are chosen according to the Lagrange polynomial basis (cf. eqs. \ref{eq:app_lagrange_weights_1} and \ref{eq:app_lagrange_weights_2} in appendix \ref{app:lagrange_interpolation}), it turns out that the general quadrature error bound in eq. \ref{eq:lagrange_scalar_error_bound} may be used in an element-wise fashion to bound the expression in eq. \ref{eq:proofsketch_lem_3_step_1}. We obtain
    \begin{align}\label{eq:proofsketch_lem_3_step_2}
        \| \mt A_{\eta, a} - \mt A_{\eta, a}^{M, q} \|_n \leq  Q^q_n \sum_{\kappa=1}^M \delta\tau_\kappa^{q + 2} \> \ee^{-\eta\tau_{\kappa-1}}; \quad\quad Q^q_n = \frac{(3 \|\H\|)^{q + 1} \|\partial_\lambda \H\|_n}{(q + 1)!}.
    \end{align}
    The next step is to choose the intervals $\delta\tau_\kappa$. The motivation behind the choice in eq. \ref{eq:tau_kappa} is to make the intervals small when $\ee^{-\eta \tau_\kappa}$ is large, and large when $\ee^{-\eta \tau_\kappa}$ is small. In this way, every term in the sum in eq. \ref{eq:proofsketch_lem_3_step_2} is almost constant, and can be bounded precisely by a constant at most $Q_n^q (2a / M)^{q + 2}$ assuming $M \geq \ee^{\eta a / (q + 2)} - 1$. This gives a better error bound than, for example, intervals of constant size. \par
    In the end, with a bit of rewriting we find
    \begin{align}
        \|(\mt U_{\eta, a}(\lambdaf, \lambdai) - \mt U_{\eta, a}^{M, q}(\lambdaf, \lambdai))\ket{n(\lambdai)}\|
        &\leq \frac{3^{q + 1}(2a)^{q + 2}}{M^{q + 1}(q + 1)!} \|\H\|_{\infty, \infty}^{q + 1} \|\partial_\lambda \H(\lambda)\|_{n, 1}
    \end{align}
    and this error can be made at most $\epsilon$ by taking $M$ as in eq. \ref{eq:M_condition}.
\end{proofsketch}
The detailed proof is deferred to appendix \ref{app:proof_summation_error}.

\subsection{Lie-Trotter-Suzuki expansion of the weighted sum}
The path-ordered exponential $\mt U_{\eta, a}^{M, q}(\lambdaf, \lambdai) = \mathcal P\exp[-\ii \int_\lambdai^\lambdaf \d\lambda \, \mt A_{\eta, a}^{M, q}(\lambda)]$ may be implemented by a $k$\textsuperscript{th}-order $r$-segment LTS product formula. These formulas follow the same description as given in section \ref{sec:LTS}, except we will be working in $\lambda$ space instead of time space. \par
The first-order single-segment formula $\tilde{\mt U}_{1, 1}(\lambdaf, \lambdai)$, for the operator 
\begin{align}
    \mt A_{\eta, a}^{M, q}(\lambda) = \sum_{\kappa = -M}^M \sum_{\alpha = 0}^q \mt C_{\kappa, \alpha}(\lambda)
\end{align}
where
\begin{align}\label{eq:C_B_def}
    \mt C_{\kappa, \alpha}(\lambda) = \ee^{-\ii \H(\lambda) \tau_{\kappa, \alpha}} \> \mt B_{\kappa, \alpha}(\lambda) \, \ee^{\ii \H(\lambda) \tau_{\kappa, \alpha}}; \quad\quad
    \mt B_{\kappa, \alpha}(\lambda) = \frac12 \delta\tau_\kappa w_{\kappa, \alpha} \, \ee^{-\eta|\tau_{\kappa, \alpha}|} \sgn\tau_{\kappa, \alpha} \> \partial_\lambda \H(\lambda)
\end{align}
(with $\mt C_{0, \alpha} = 0$), is given by
\begin{align}\label{eq:tildeU1,1_CD}
    \tilde{\mt U}_{1, 1}(\lambdaf, \lambdai) = \bigg( \prod_{\kappa=-M}^M \prod_{\alpha = 0}^q \exp [-\ii \mt C_{\kappa, \alpha}(\lambdai + \delta\lambda / 2) \delta\lambda / 2] \bigg) \bigg( \prod_{\kappa=M}^{-M} \prod_{\alpha = q}^0 \exp [-\ii \mt C_{\kappa, \alpha}(\lambdai + \delta\lambda / 2) \delta\lambda / 2] \bigg).
\end{align}
where $\delta\lambda = \lambdaf - \lambdai$. Here, the operator exponentials in the product take the form
\begin{align}\label{eq:exp_iC_decomp}
    \exp[-\ii \mt C_{\kappa, \alpha}(\lambda) \delta\lambda] = \exp[-\ii \H(\lambda) \tau_{\kappa, \alpha}] \> \exp\!\bigg[ \!-\!\frac\i2 \ee^{-\eta|\tau_{\kappa, \alpha}|} \sgn\tau_{\kappa, \alpha} w_{\kappa, \alpha} \> \partial_\lambda \H(\lambda) \> \delta\tau_\kappa \> \delta\lambda \bigg] \exp[\ii \H(\lambda) \tau_{\kappa, \alpha}].
\end{align}
Higher-order, multi-segment formulas are constructed as in eqs. \ref{eq:LTS_order_k+1} and \ref{eq:LTS_U_k_r}. In the same way, we can use lemma \ref{lem:Wiebe} to put a bound on the number of path-independent operator exponentials required to achieve a certain precision. We apply this result to our case by giving an explicit expression for $\Lambda$ corresponding to the ordered exponential $\U_{\eta, a}^{M, q}$. If the spectral distance between $\U_{\eta, a}^{M, q}$ and its LTS decomposition $\tilde\U_{k, r}$ is at most $\epsilon$,  then the euclidean state distance $\|(\mt U_{\eta, a}^{M, q}(\lambdaf, \lambdai) - \tilde\U_{k, r}(\lambdaf, \lambdai)) \ket{n(\lambdai)}\|$ is also at most $\epsilon$. \par
To use lemma \ref{lem:Wiebe}, we require bounds on the operator norms of the higher-order derivatives of the $\mt C_{\kappa, \alpha}$ operators. For the case of interpolating hamiltonians $\H(\lambda) = \Hi + f(\lambda) \Hp$, these norms take on a simple form, since
\begin{align}
    \bigg\| \frac{\partial^p}{\partial\lambda^p} \big( \ee^{-\ii \H(\lambda) \tau} \dH \ee^{-\ii \H(\lambda) \tau} \big) \bigg\| = \bigg\| \frac{\partial^{p + 1} f(\lambda)}{\partial\lambda^{p + 1}} \> \Hp \bigg\| = \|\partial_\lambda^{p + 1} \H(\lambda)\|;
\end{align}
however, for general hamiltonians, the higher-order derivatives of $\mt C_{\kappa, \alpha}(\lambda)$ are less straightforward and their norms can grow rapidly with $p$, as a result of compounding product rules.
Since there can be a significant discrepancy in the the scaling of the derivatives between interpolating case $\H(\lambda) = \Hi + f(\lambda) \Hp$ and the general case, and better bounds on the norm of $\|(\partial^p / \partial\lambda^p) \mt C_{\kappa, \alpha}(\lambda)\|$ require more specifics about the definition of $\H(\lambda)$, we will stick to the interpolating case for the rest of the section. A condition on the number $r$, and thereby the number of exponentials required in our gate-based counterdiabatic driving algorithm, is then given by the following lemma.

\begin{lemma}\label{lem:LTS_error}
    Let $\H(\lambda) = \Hi + f(\lambda) \Hp$ for hermitian $\Hi$ and $\Hp$ and a scalar function $f(\lambda)$ defined on the interval $[\lambdai, \lambdaf]$ that is $2k + 1$ times differentiable. Let $\U_{\eta, a}^{M, q}(\lambdaf, \lambdai)$, $\{\tau_\kappa\}$, $\{\tau_{\kappa, \alpha}\}$ and $\{w_{\kappa, \alpha}\}$ be as in lemma \ref{lem:summation_error}, and let $\tilde\U_{k, r}(\lambdaf, \lambdai)$ be the $k$\textsuperscript{th}-order $r$-segment LTS decomposition of $\U_{\eta, a}^{M, q}(\lambdaf, \lambdai)$ as in eq. \ref{eq:LTS_U_k_r}. Let the conditions of lemma \ref{lem:summation_error} be satisfied. Suppose
    \begin{align}\label{eq:Lambda_CD}
        \max_{1 \leq p \leq 2k + 1} \bigg[ \bigg( \frac{2(1 - \ee^{-\eta a})}\eta \> \|\partial_\lambda^p \H\|_{\infty, \infty} \bigg)^{1 / p} \, \bigg] \leq \tilde\Lambda.
    \end{align}
    If $\|\dH\|_{n, 1} \geq 3^{-(q + 1)} \frac\eta{1 - \ee^{-\eta a}}$ and $\epsilon \leq \min\{(9/10)(5/3)^k \> \tilde\Lambda \> \delta\lambda, 1\}$, then
    \begin{align}
        \|\U_{\eta, a}^{M, q}(\lambdaf, \lambdai) - \tilde\U_{k, r}(\lambdaf, \lambdai)\| \leq \epsilon
    \end{align}
    provided that
    \begin{align}\label{eq:r_condition}
        r \geq 5k \> \tilde\Lambda \> \delta\lambda \> \bigg( \frac53 \bigg)^k \bigg( \frac{\tilde\Lambda \delta\lambda}\epsilon \bigg)^{1/2k}.
    \end{align}
\end{lemma}

\begin{proofsketch}
    We identify $\tilde\Lambda$ with the quantity $\Lambda$ from lemma \ref{lem:Wiebe}. We therefore need to bound the sum
    \begin{align}\label{eq:proofsketch_lem_4_step_1}
        \sum_{\kappa=-M}^M \sum_{\alpha = 0}^q \bigg\| \frac{\partial^p}{\partial \lambda^p} \mt C_\kappa(\lambda) \bigg\| &= \frac12 \|\partial_\lambda^{p+1} \H(\lambda)\| \sum_{\kappa=-M}^M \sum_{\alpha = 0}^q \delta\tau_\kappa w_{\kappa, \alpha} \, \ee^{-\eta|\tau_{\kappa, \alpha}|}.
    \end{align}
    Since the right-hand side of eq. \ref{eq:proofsketch_lem_4_step_1} is a discrete approximation of the integral $\int_{-a}^a \d\tau \, \ee^{-\eta |\tau|}$ with Lagrange weights, we can use the same bounds as in the proof of lemma \ref{lem:summation_error}. This is a straightforward calculation that leads to
    \begin{align}
        \sum_{\kappa=-M}^M \sum_{\alpha = 0}^q \bigg\| \frac{\partial^p}{\partial \lambda^p} \mt C_\kappa(\lambda) \bigg\|  \leq \frac{2(1 - \ee^{-\eta a})}\eta \> \|\partial_\lambda^{p + 1} \H(\lambda)\|.
    \end{align}
    Therefore the condition on $\tilde\Lambda$ in eq. \ref{eq:Lambda_CD} suffices to satisfy the requirements of lemma \ref{lem:Wiebe}, and the result follows.
\end{proofsketch}
The calculation is worked out in appendix \ref{app:proof_LTS_error}. The assumption on $\|\dH\|_{n, 1}$ is justified since $3^{-(q + 1)}$ can be made arbitrarily small and we work in the regime where $\eta$ is small and $a$ is large (note $\frac\eta{1 - \ee^{-\eta a}}$ approaches zero in the limit $a \to \infty$, $\eta \to 0$).

\subsection{Gate complexity}
Finally, we turn to the quantum gate complxity of the gate-based CD algorithm. In order to bound this complexity, we need to associate with each path-independent operator exponential in the product formula $\tilde\U_{k, r}(\lambdaf, \lambdai)$ a gate complexity and sum over all operator exponentials. In the case of AQC, all these operator exponentials were assumed to be simulatable with a single multi-qubit rotation gate, and unit cost was associated with each simulation. This is not the case for CD, since the constituents of the LTS formulae are exponentials of $\H(\lambda)$ and $\dH(\lambda)$. Simulation costs for these exponentials in terms of quantum gates are provided through established time-independent hamiltonian simulation methods. We use a technique known as qubitisation \cite{Low2019b} for its optimality in the relevant parameters. In short, if an operator $\H$ is given as a linear combination of unitaries\footnote{The qubitisation subroutine requires that the hamiltonian be supplied as a LCU or as a sparse matrix with oracle entry access. Since the former is more natural and more common form for most physics settings (in particular many-body problems), we go with the LCU formulation here.} $\H = \sum_{i = 1}^\ell \beta_i \mt V_i$, qubitisation provides a routine that simulates $\ee^{\pm \ii \H \theta}$ to error $\epsilon$ in spectral norm using $O(\|\vcg\beta\|_1 \theta + \log 1 / \epsilon)$ queries to an oracle that provides access to the hamiltonian, where $\|\vcg\beta\|_1 = \sum_j |\beta_j|$. Since this oracle can be implemented with $O(\ell)$ elementary gates \cite{Shende2006, Childs2012, Low2019b}, the simulation can be done with $O([\|\vcg\beta\|_1 \theta + \log 1 / \epsilon] \ell)$ gates.\footnote{Since the qubitisation method is rather technical and its details are out of the scope of this work, we will use it as a black-box subroutine in the rest of the paper. Nevertheless, the method is well known, and for readers interested in implementing our algorithm on quantum hardware or simulators, libraries exist that implement this subroutine for out-of-the-box usage. See for example ref. \citenum{PennylaneQubitization}.} \par
With a clear definition of gate complexity, we can now state the main theorem of this paper, establishing a gate complexity bound for our gate-based counterdiabatic driving algorithm. 
\begin{theorem}[gate-based CD]\label{thm:gate_based_CD}
    Suppose $\H(\lambda) = \Hi + f(\lambda) \Hp = \sum_{i = 1}^\ell \beta_i(\lambda) \mt V_i$ with hermitian $\Hi$ and $\Hp$ and a scalar function $f(\lambda)$ that is $2k + 1$ times differentiable on the interval $[\lambdai, \lambdaf]$, $\beta_i(\lambda) \in \R$ and $\mt V_i$ unitary. Let $\{\ket{n(\lambda)}\}$ be the set of its instantaneous eigenstates. Suppose the spectrum of $\H(\lambda)$ around an eigenstate $\ket{n(\lambda)}$ has a gap of at least $\Delta_n$ on the entire interval. Let $\U(\lambdaf, \lambdai) = \sum_n \ket{n(\lambdaf)}\bra{n(\lambdai)}$. Define $\gamma_p = (\sqrt2 \Delta_n^{-3 / 2} \|\dH\|_{n, 1}^{1 / 2} \> \|\partial_\lambda^p \H\|_{\infty, \infty})^{1 / p}$, $p^* = \argmax_{1 \leq p \leq 2k + 1} (\epsilon^{-1 / 2p} \gamma_p)$ and $\tilde\Lambda = \epsilon^{-1 / 2p^*} \gamma_{p^*}$. Furthermore, let $\epsilon > 0$, $q \in \N_{>0}$ and suppose that
    \begin{enumerate}[label=(\alph*)]
        \item $\epsilon \leq \min\{[(9 / 10)(5 / 3)^k \gamma_{p^*} (\lambdaf - \lambdai)]^{1 - \frac1{2p^* + 1}}, 1\}$;
        \item $\|\dH\|_{n, 1} \geq 3^{-2 (q + 1) / 3} \Delta_n \epsilon^{1 / 3}$;
        \item $\min_{\lambda \in [\lambdai, \lambdaf]} \|\H(\lambda)\| \geq \Delta_n^{3 / 2} \epsilon^{1 / 2} \|\dH\|_{n, 1}^{-1 / 2}$.
    \end{enumerate}
    Then there exists a quantum algorithm $\tilde\U(\lambdaf, \lambdai)$ such that
    \begin{align}\label{eq:final_CD_error}
        \|(\U(\lambdaf, \lambdai) - \tilde\U(\lambdaf, \lambdai))\ket{n(\lambdai)}\| \leq O(\epsilon)
    \end{align}
    which can be implemented with
    \begin{align}\label{eq:final_CD_gate_complexity_bound}
        \tilde O\Bigg( \bigg( \frac{25}3 \bigg)^k k \> (\Lambda \> \delta\lambda)^{1 + \frac1{2k}} \frac{\|\dH\|_{n, 1}^{1 + \frac1{4k} + \frac3{2(q + 1)}}}{\epsilon^{1 + \frac3{4k} + \frac3{2(q + 1)}} \Delta_n^{3 + \frac3{4k} + \frac3{2(q + 1)}}} \|\vcg\beta\|_{1, \infty} \Bigg)
    \end{align}
    quantum gates, where $\|\vcg\beta\|_{1, \infty} = \max_{\lambda \in [\lambdai, \lambdaf]} \sum_{i = 1}^\ell |\beta_i(\lambda)|$ and $\Lambda = \max_{1 \leq p \leq 2k + 1} \|\partial_\lambda^p \H\|_{\infty, \infty}^{1 / p}$.
\end{theorem}

\begin{proofsketch}
    Let $\tilde\U_{k, r}$ be as in lemma \ref{lem:LTS_error}; assumptions (a)--(c) serve to fulfill the conditions of lemmas \ref{lem:summation_error} and \ref{lem:LTS_error} so that, by a triangle inequality, $\|(\U(\lambdaf, \lambdai) - \tilde\U_{k, r}(\lambdaf, \lambdai))\ket{n(\lambdai)}\| \in O(\epsilon)$ provided that $\eta$ and $a$ are set as in lemma \ref{lem:regularised_truncated_unitary_error}. To finalise the proof, we must determine how $\tilde\U_{k, r}$ is decomposed into elementary quantum gates to precision $\epsilon$ (in spectral norm) and how many are needed. Since $\tilde\U_{k, r}$ is a sequence of alternating exponentials of the form $\ee^{-\ii \H \theta}$ and $\ee^{-\ii \dH \theta}$, we only need to count the number of such exponentials and apply the standard qubitisation result mentioned earlier. Carrying out this counting and identifying the dominant terms yields
    \begin{align}
        \text{gate complexity} \in \tilde O\big( \ell \> 5^{k - 1} r \kern0.03em M (q + 1) \big)
    \end{align}
    where $M$ and $q$ are the number of subintervals and interpolation points in the integral approximation, as before. We note that the dependence on the precision $\epsilon$ is logarithmic and has been absorbed into the $\tilde O(\kern0.13em\cdots)$ notation. Inserting the values for $M$ from lemma \ref{lem:summation_error} (eq. \ref{eq:M_condition}) and $r$ from lemma \ref{lem:LTS_error} (eq. \ref{eq:r_condition}), recognising that $\tilde\Lambda \leq 2\eta^{-1} \Lambda$ and subsequently inserting $\eta$ and $a$ from lemma \ref{lem:regularised_truncated_unitary_error} (eq. \ref{eq:eta_a_conditions}) then lead to the final result.
\end{proofsketch}
We provide the complete proof in appendix \ref{app:proof_gate_based_CD}.

\section{Comparison to gate-based adiabatic computing}\label{sec:gate_based_AQC}
We now return to the matter of comparing counterdiabatic driving to adiabatic computing. The complexity of AQC is typically quantified by the physical evolution time $T$ which sets the speed at which $\lambda(t)$ is varied. However, in the previous sections we have quantified the complexity of CD in terms of the number of digital quantum operations (quantum gates); as such, physical time, with which the performance of adiabatic computing is typically quantified, is insufficient here since it is not a meaningful quantity in the description of gate-based CD and therefore leads to an apples-to-oranges comparison. Instead, we will describe the complexity of AQC in the same gate-based fashion, counting the number of quantum gates with which gate-based AQC can be implemented. In this context, the standard adiabatic theorems which state bounds on the physical time are still important, since in the end, the number of quantum gates that is needed for gate-based AQC is mostly determined by the physical time. \par
Several versions of the adiabatic theorem exist. Nonrigorous, ``folk'' versions typically estimate $T \in \Omega(\Delta_n^{-2})$; but since we conducted a rigorous complexity analysis of CD, such statements do not suffice for a comparison. Rigorous theorems typically yield a cubic or near-quadratic worst-case scaling in $\Delta_n^{-1}$, though a quadratic dependence may be obtained \cite{Elgart2012} if it is assumed that $\H(\lambda)$ belongs to a Gevrey class\footnote{A hamiltonian $\H(\lambda)$ belongs to a Gevrey class $G^\alpha$ if $\dH \neq 0 \; \forall \lambda \in [\lambdai, \lambdaf]$ and there exist constants $C, D > 0$ such that for all $p > 1$, $\max_{\lambda \in [\lambdai, \lambdaf]} \|\partial_\lambda^p \H(\lambda)\| \leq C D^p k^{\alpha k}$.}. A general result, and, to the best of our knowledge, the tightest bound in terms of the minimum gap $\Delta_n$ is due to Reichardt \cite{Reichardt2004}. We quote this result here.

\begin{lemma}[adapted from Reichardt \cite{Reichardt2004}]\label{lem:Reichardt}
    Let $T > 0$, and let $\lambda(\cdot) : \R \to \R$ be a function such that $\lambda(0) = \lambdai$ and $\lambda(T) = \lambdaf$. Suppose $\H(\lambda)$ is hermitian, gapped and $K + 1$ times differentiable in $\lambda$ on the interval $[\lambdai, \lambdaf]$, with $p \geq 1$. Let $\U(T, 0) = \mathcal T\exp[-\ii\int_0^T \d t \, \H(\lambda(t))]$. Let $\ket{n(\lambdai)}$ be the the $n$-th instantaneous eigenstate of $\H(\lambdai)$, and $\ket{n(\lambdaf)} = \lim_{T\to\infty} \U(T, 0)\ket{n(\lambdai)}$ (i.e. the $n$-th eigenstate of $\H(\lambdaf)$), and assume $\|\partial_\lambda^p \H(\lambda)\| \leq \Gamma$ for all $s \leq K + 1$. Then
    \begin{align}
        |\bra{n(\lambdaf)} \U(T, 0) \ket{n(\lambdai)}|^2 \geq 1 - \epsilon^2
    \end{align}
    provided that
    \begin{align}\label{eq:AQC_time_complexity}
        T \geq \Omega(\epsilon^{-1 / K} \Delta_n^{-(2 + 1 / K)} \Gamma^{1 + 1 / K})
    \end{align}
    where $\Delta_n$ is the minimum spectral gap around $\ket{n(\lambda)}$ on the interval $[\lambdai, \lambdaf]$.
\end{lemma}


We proceed to put a bound on the gate complexity of gate-based AQC by combining lemmas \ref{lem:Reichardt} and \ref{lem:Wiebe}. Lemma \ref{lem:Wiebe} gives a bound on the number of operator exponentials with which a time evolution in a time interval of given length can be simulated by a LTS decomposition (cf. section \ref{sec:LTS}). If we assume the hamiltonian is given as a linear combination of unitaries $\H(t) = \sum_{i = 1}^\ell \beta_i(t) \mt V_i$, every exponential is a multi-qubit rotation gate; if required, such gates can be decomposed into a number of elementary gates (single-qubit rotations, CNOT gates and the like) that depends only on the locality of the respective unitary $\mt V_i$ (i.e. the number of qubit it acts on nontrivially). We will consider the maximum locality of these terms as a constant, and as such, in big-$O$ notation, we equate the number of elementary gates to the number of (time-independent) operator exponentials. The following then gives a bound on the number of operator exponentials to implement gate-based AQC.

\begin{proposition}[gate-based AQC]\label{prop:gate_based_AQC}
    Let $T > 0$ as in lemma \ref{lem:Reichardt}, and let $\lambda(t)$ be such that $\lambda(0) = \lambdai$ and $\lambda(T) = \lambdaf$. Suppose $\H(\lambda(t)) = \sum_{i = 1}^\ell \beta_i(\lambda(t)) \mt V_i$ is hermitian and $2k + 1$ times differentiable in $t$ on the interval $[0, T]$, with $\beta_i(\lambda(t)) \in \R$ and $\mt V_i$ unitary. Let $\{\ket{n(\lambda(t))}\}$ be the set of its instantaneous eigenstates. Suppose the spectrum of $\H(\lambda(t))$ around an eigenstate $\ket{n(\lambda(t))}$ has a gap of at least $\Delta_n$ on the entire interval. Let $\U(\lambdaf, \lambdai) = \sum_n \ket{n(\lambdaf)}\bra{n(\lambdai)}$. Assume $\|\partial_\lambda^p \H(\lambda)\| \leq \Gamma$ for all $p \leq 2k + 1$ and $T \in \Theta(\epsilon^{-1 / 2k} \Delta_n^{-(2 + 1 / 2k)} \Gamma^{1 + 1 / 2k})$. Then there exists a gate-based quantum algorithm $\tilde\U(T, 0)$ such that
    \begin{align}\label{eq:AQC_error}
        \|(\mt U(\lambdaf, \lambdai) - \tilde\U(T, 0)) \ket{n(\lambdai)}\| \leq O(\epsilon)
    \end{align}
    which can be implemented with 
    \begin{align}\label{eq:AQC_complexity}
        O\Bigg( \ell \bigg( \frac{25}3 \bigg)^k k \> \Lambda^{1 + \frac1{2k}} \frac{\Gamma^{(1 + \frac1{2k})^2}}{\epsilon^{\frac1{2k}(2 + \frac1{2k})} \Delta_n^{(2 + \frac1{2k})(1 + \frac1{2k})}} \Bigg)
    \end{align}
    quantum gates, where $\Lambda = \max_{0 \leq p \leq 2k} \|\partial_t^p \vcg\beta\|_{1, \infty}^{1 / (p + 1)}$ and $\|\partial_t^p \vcg\beta\|_{1, \infty} = \max_{t \in [0, T]}\sum_{i = 1}^\ell |\partial_t^p \beta_i|$.
\end{proposition}
\begin{proof}
    Let $\U(T, 0) = \mathcal T \exp[-\ii \int_0^T \d t \, \H(t)]$ and identify $\tilde\U(T, 0)$ as a $k$\textsuperscript{th}-order $r$-segment LTS decomposition of $\U(T, 0)$. Observe that a first-order single-segment LTS decomposition $\tilde\U_{1, 1}$ of $\U(T, 0)$ (eq. \ref{eq:tilde_U_1,1_definition}) contains $2\ell$ operator exponentials, and that $\tilde\U_{k, r}$ is a product of $2\ell \> 5^{k - 1} r$ operator exponentials. Lemma \ref{lem:Reichardt} gives the conditions under which $\|(\U(\lambdaf, \lambdai) - \U(T, 0)) \ket{n(\lambdai)}\| \leq O(\epsilon)$; we use the bound on $r$ from lemma \ref{lem:Wiebe}, which assures that $\|(\U(T, 0) - \tilde\U(T, 0)) \ket{n(\lambdai)}\| \leq O(\epsilon)$ so that, by a triangle inequality, eq. \ref{eq:AQC_error} holds. To use this bound, we first notice that the choice $\Lambda = \max_{0 \leq p \leq 2k} \|\partial_t^p \vcg\beta\|_{1, \infty}^{1 / (p + 1)}$ is sufficient to fulfill the requirement of eq. \ref{eq:Wiebe_Lambda} in lemma \ref{lem:Wiebe}. From lemma \ref{lem:Reichardt} we then insert $T$ to obtain that
    \begin{align}
        & O\bigg( \ell \bigg( \frac{25}3 \bigg)^k k \> \Lambda^{1 + \frac1{2k}} T^{1 + \frac1{2k}} \epsilon^{-\frac1{2k}} \bigg)
        = O\Bigg( \ell \bigg( \frac{25}3 \bigg)^k k \> \Lambda^{1 + \frac1{2k}} \frac{\Gamma^{(1 + \frac1{2k})^2}}{\epsilon^{\frac1{2k}(2 + \frac1{2k})} \Delta_n^{(2 + \frac1{2k})(1 + \frac1{2k})}} \Bigg)
    \end{align}
    quantum gates are sufficient to guarantee an overall error of at most $O(\epsilon)$.
\end{proof}

In short, we find that the physical time $T \in \Theta(\epsilon^{-o(1)} \Delta_n^{-(2 + o(1))})$ leads to a similar gate complexity upper bound of $O(\epsilon^{-o(1)} \Delta_n^{-(2 + o(1))})$. \par
Besides the bound by Reichardt and the subsequent gate complexity, another result that deserves a mention is that by Jansen et al. \cite{Jansen2007}. Their bound is $T \geq \Omega\big( \epsilon^{-1} \int_\lambdai^\lambdaf \d\lambda \, \|\dH(\lambda)\|^2 / \omega_n(\lambda)^3\big)$ for hamiltonians that are twice differentiable, where $\omega_n(\lambda)$ is the instantaneous gap around the eigenstate $\ket{n(\lambda)}$ of $\H(\lambda)$. At first glance, this seems equivalent to Reichardt's result for $K = 1$. However, the authors point out that, at least for unstructured (Grover) search, the result should not be viewed as an inverse cubic dependence on the mimimum gap. They show that instead, $\int_\lambdai^\lambdaf \d\lambda \,\omega_n(\lambda)^{-3} \in \Theta(\Delta_n^{-2})$, and that further improvement is achieved by tuning the interpolation function $f(\lambda)$ to the instantaneous gap, which leads to a linear scaling in $\Delta_n^{-1}$. Similar results were obtained for the quantum linear systems problem (QLSP) by Cunningham et al. \cite{Cunningham2024}. Looking at proposition \ref{prop:gate_based_AQC}, we would therefore also find a quadratic or linear inverse gap scaling in the number of quantum gates. \par

Comparing the $\tilde O(\epsilon^{-(1 + o(1))} \Delta_n^{-(3 + o(1))})$ bound from theorem \ref{thm:gate_based_CD} for CD to the $\tilde O(\epsilon^{-o(1)} \Delta_n^{-(2 + o(1))})$ from proposition \ref{prop:gate_based_AQC} for AQC, it appears that there is room for improvement in the analysis of gate-based CD. One such improvement can be easily spotted: in the proof of lemma \ref{lem:regularised_truncated_unitary_error}, we determined $\eta$ from the condition $2 \eta^2 \int_\lambdai^\lambdaf \d\lambda \, \frac{\|\dH(\lambda)\|}{\omega_n^3(\lambda)} \leq 2 \eta^2 \|\dH\|_{\infty, 1} \Delta_n^{-3} \leq \epsilon$. However, in the same vein as the argument put forward by Jansen et al. \cite{Jansen2007}, the integral may in certain cases be reduced to $\Theta(\Delta_n^{-2})$, which leads to a linear gap dependence $\eta \in \Theta(\epsilon^{1 / 2} \Delta_n)$. Since the gate complexity of CD depends at most quadratically on $\eta^{-1}$ (up to logarithmic factors), we obtain an (almost) inverse quadratic overall gap scaling instead of inverse cubic. Improving this to linear for problems like unstructured search is however less straightforward, and likely requires a variable $\eta(\lambda)$ throughout the driving, which may be tuned to the instantaneous gap, instead of a constant $\eta$. In principle, one could adopt optimisation schemes to find good cutoff functions $\eta(\lambda)$, in a way similar to optimisation of scheduling functions \cite{Cote2023}. However, given the difficulty of analysing optimisation methods, this would add a layer of uncertainty regarding the computational complexity of the complete procedure.

\section{Conclusion}\label{sec:conclusion_outlook}
In this work, we have presented what is, to the best of our knowledge, the first fully gate-based quantum algorithm for counterdiabatic driving. This algorithm is constructed from the regularised truncated adiabatic gauge potential (eq. \ref{eq:regularised_truncated_AGP}). By discretising the integral form of this AGP approximation, it is fed into a Lie-Trotter-Suzuki formula to produce a gate-based algorithm that may be run on a quantum computer. We have shown that this algorithm, starting from an initial eigenstate $\ket{n(\lambdai)}$, can be run with $\tilde O(\epsilon^{-(1 + o(1))} \Delta_n^{-(3 + o(1))}))$ quantum gates in the worst case in order to achieve a fidelity at least $1 - \epsilon^2$ with the target eigenstate $\ket{n(\lambdaf)}$ (theorem \ref{thm:gate_based_CD}). Here, $\Delta_n$ is the minimum energy gap around the instantaneous eigenstate $\ket{n(\lambda)}$. The $o(1)$ scaling is an inverse linear dependence on the order $k$ of the LTS formula (see eqs. \ref{eq:tildeU1,1_CD} and \ref{eq:LTS_order_k+1}) and the degree $q$ of the Lagrange interpolation polynomial used in the discretisation of the integral form. We remark that $q$ can be made large cheaply using purely classical precomputation of the quadrature weights. \par
This scaling is almost equivalent to the $O(\Delta^{-(3 + o(1))} \epsilon^{-(1 + o(1))})$ gate complexity bound that follows from applying LTS simulation to the worst-case result of Jansen et al. \cite{Jansen2007} (up to logarithmic factors), but is polynomially worse than the $O(\Delta^{-(2 + o(1))} \epsilon^{-o(1)})$ bound that results in a similary way from the scaling given by Reichardt \cite{Reichardt2004}. However, an inverse quadratic gap scaling $\tilde O(\Delta^{-(2 + o(1))} \epsilon^{-1 + o(1)})$ may be achieved with our gate-based CD method in certain cases such as unstructured (Grover) search and the QLSP problem. \par
All in all, CD remains an interesting approach for quantum state preparation. After all, there may exist more efficient gate-based CD algorithms or better complexity bounds. And while CD cannot be generally faster than AQC, at least in terms of gap scaling -- this would violate Grover optimality -- our work opens up a new formalism for finding gate complexity upper bounds based on an approximation of the AGP. \par
Furthermore, CD can still be valuable in settings where quantum resources are scarce. For example, in noisy setups where the ability to work on small timescales is paramount to countering noise, adiabaticity is not available; achieving satisfactory fidelities then necessitates the use of some kind of classically precomputed shortcut field. We also remark that the algorithm presented in this work is intended as a fault-tolerant routine, whose implementation may remain out of reach for the near future, especially for settings with small minimum gaps which lead to deep circuits. \par
Lastly, we remark that, in order to obtain the presented gate complexity of CD, it turned out crucial to tailor the AGP approximation to the eigenstate of interest, through an appropriately chosen nonzero gap cutoff. This insight that it is unnecessary to suppress all transitions in the spectrum allowed us to reduce the complexity down from the dimensionality of the operator space to a scaling in the minimum gap around that specific eigenstate. While this idea has been recognised \cite{Demirplak2008, Morawetz2024}, it was not taken into account in the currently most prevalent algorithmic CD approaches \cite{Sels2017, Claeys2019, Bhattacharjee2023, Takahashi2024}. As such, we would like to stress its importance once again, and hope to see it replicated in future work on counterdiabatic driving. \par

\section*{Acknowledgments}
The author thanks Kareljan Schoutens for his suggestion of the differential equation ansatz in eq. \ref{eq:tau_diff_eq}. Additional thanks go to Ronald de Wolf, Jan St\v{r}ele\v{c}ek and Takuya Hatomura for insightful discussions. This work was supported by the Dutch Ministry of Economic Affairs and Climate Policy (EZK), through the Quantum Delta NL programme.

\bibliographystyle{unsrt}
\bibliography{biblio}

\appendix

\section{Proof of lemma \ref{lem:regularised_truncated_unitary_error}}\label{app:proof_reg_trunc_unit_err}
\begin{proof}
We first bound the unitary error by an error in the AGP:
\begin{align}\label{eq:regularised_truncated_unitary_AGP_error}
    \|(\mt U(\lambdaf, \lambdai) - \mt U_{\eta, a}(\lambdaf, \lambdai)) \ket{n(\lambdai)}\|
    &= \|(\mt U_{\eta, a}\t(\lambdaf, \lambdai)\mt U(\lambdaf, \lambdai) - \mt I) \ket{n(\lambdai)}\| \nonumber \\
    &= \bigg\| \int_\lambdai^\lambdaf \d\lambda \, \frac\d{\d\lambda} (\mt U_{\eta, a}\t(\lambda, \lambdai)\mt U(\lambda, \lambdai)) \ket{n(\lambdai)} \bigg\| \nonumber \\
    &= \bigg\| \int_\lambdai^\lambdaf \d\lambda \, \mt U_{\eta, a}\t(\lambda, \lambdai) \> [\mt A(\lambda) - \mt A_{\eta, a}(\lambda)] \> \mt U(\lambda, \lambdai) \ket{n(\lambdai)} \bigg\| \nonumber \\
    &\leq \int_\lambdai^\lambdaf \d\lambda \, \|(\mt A(\lambda) - \mt A_{\eta, a}(\lambda)) \ket{n(\lambda)}\|
\end{align}
where the last inequality follows from a triangle inequality on the integral. To evaluate the AGP error, we make use of the energy eigenbasis expansion (omitting the $\lambda$ argument for brevity and using that $\braket k{\partial_\lambda k} = 0$)
\begin{align} \label{eq:time_translated_dlambdaH_energy_eigenbasis}
    \ee^{-\ii\H\tau} \dH \> \ee^{\ii\H\tau}
    &= \sum_k \partial_\lambda E_k \ket k\bra k + \sum_{k; \, m\neq k} \ee^{-\ii (E_m - E_k) \tau} \ket m \bra m \dH \ket k \bra k.
\end{align}
Since $\ee^{-\eta|\tau|}$ is symmetric in $\tau$, it follows that\footnote{We abuse notation in eqs. \ref{eq:abuse_nonconvergent_integral_1} and \ref{eq:abuse_nonconvergent_integral_2} since $\int_{-\infty}^\infty \d\tau \> \sgn\tau$ does not converge. The implied meaning is $\lim_{\mu\to0} \int_{-\infty}^\infty \d\tau \, \ee^{-\mu|\tau|} \sgn\tau$.}
\begin{align}\label{eq:abuse_nonconvergent_integral_1}
    \sum_k \frac12 \int_{-\infty}^\infty \d\tau \sgn(\tau) \> (1 - \ee^{-\eta|\tau|}\mathds 1_{\tau\in[-a, a]}) \> \partial_\lambda E_k \ketbra k = 0.
\end{align}
Furthermore, for any eigenstate $\ket n$,
\begin{align}\label{eq:abuse_nonconvergent_integral_2}
    & \sum_{\substack{k \\ m\neq k}} \frac12 \int_{-\infty}^\infty \d\tau \sgn(\tau) \> (1 - \ee^{-\eta|\tau|}\mathds 1_{\tau\in[-a, a]}) \> \ee^{-\ii (E_m - E_k) \tau} \ket m \bra m \dH \ket k \braket kn \nonumber \\
    &= {-\ii} \sum_{m\neq n} c_{\eta, a}(\omega_{mn}) \bra m \dH \ket n \> \ket m
\end{align}
where
\begin{align}
    c_{\eta, a}(\omega) := \frac1{\omega} - \frac{\omega}{\omega^2 + \eta^2} + \ee^{-a \> \eta} \frac{\cos(a\omega) \> \omega + \sin(a\omega) \> \eta}{\omega^2 + \eta^2}.
\end{align}
Observe that
\begin{align}
    |c_{\eta, a}(\omega)| \leq g_{\eta, a}(\omega) := \frac{\eta^2}{|\omega|^3} + \ee^{-\eta a} \bigg( \frac1{|\omega|} + \frac1{\eta} \bigg)
\end{align}
and that $g_{\eta, a}(\omega)$ is a decreasing function of $|\omega|$; therefore  we may give the upper bound
\begin{align}\label{eq:infidelity_norm_FT_sgn_r}
    \int_\lambdai^\lambdaf \d\lambda \, \|(\mt A(\lambda) - \mt A_\eta(\lambda)) \ket{n(\lambda)}\|
    & = \int_\lambdai^\lambdaf \d\lambda \, \sqrt{\sum_{m\neq n} |c_{\eta, a}(\omega_{mn}(\lambda))|^2 |\bra{n(\lambda)}\partial_\lambda\H(\lambda)\ket{m(\lambda)}|^2} \nonumber \\
    &\leq \int_\lambdai^\lambdaf \d\lambda \, \max_m g_{\eta, a}(\omega_{mn}(\lambda)) \sqrt{\sum_{m\neq n} |\bra{n(\lambda)}\partial_\lambda\H(\lambda)\ket{m(\lambda)}|^2} \nonumber \\
    &\leq \int_\lambdai^\lambdaf \d\lambda \, \max_m g_{\eta, a}(\omega_{mn}(\lambda)) \|\dH(\lambda)\| \nonumber \\
    &\leq \|\dH\|_{n, 1} g_{\eta, a}(\min_m \min_\lambda \omega_{mn}(\lambda)) \nonumber \\
    &= \|\dH\|_{n, 1} g_{\eta, a}(\Delta_n).
\end{align}
The requirement that this error be at most $\epsilon$ is then satisfied by setting
\begin{align}
    \eta = \Delta_n^{3/2} (\epsilon / 2)^{1/2} \|\dH\|_{n, 1}^{-1/2}
    \quad \Rightarrow \quad 
    \frac{\eta^2}{\Delta_n^3} \|\dH\|_{n, 1} \leq \frac\epsilon2
\end{align}
and
\begin{align}
    a = \frac1\eta \log\bigg( \frac{\Delta_n + \eta}{\Delta_n\eta\epsilon / 2} \|\dH\|_{n, 1} \bigg)
    \quad \Rightarrow \quad 
    \ee^{-\eta a} \bigg( \frac1{\Delta_n} + \frac1{\eta} \bigg) \|\dH\|_{n, 1} \leq \frac\epsilon2
\end{align}
which proves the lemma.
\end{proof}

\section{Lagrange interpolation and scalar quadrature}\label{app:lagrange_interpolation}
Scalar functions may be approximated as weighted sums through Lagrange interpolation. The idea of this interpolation method is to approximate a function $f(x)$ in some interval $[a, b]$ by some polynomial $p_q(\tau)$ of degree at most $q$ such that the interpolation condition
\begin{align}\label{eq:lagrange_interpolation_condition}
    f(x_\alpha) = p_q(x_\alpha), \quad x_0, \cdots, x_q \in [a, b]
\end{align}
is satisfied for a set of $q + 1$ points in $[a, b]$. It can be shown that such a polynomial is unique, and is given by
\begin{align}
    p_q(x) = \sum_{\alpha = 0}^q f(x_\alpha) l_\alpha(x)
\end{align}
where
\begin{align}\label{eq:app_lagrange_weights_1}
    l_\alpha(x) = \prod_{\substack{0\leq\beta\leq q \\ \beta\neq\alpha}} \frac{x - x_\beta}{x_\alpha - x_\beta}
\end{align}
are the Lagrange basis polynomials. It is straightforward to check that $l_\alpha(x_\beta) = \delta_{\alpha\beta}$, which also verifies the interpolation condition (eq. \ref{eq:lagrange_interpolation_condition}). Clearly, if $f$ itself is a polynomial of degree at most $q$, then $p_q = f$ since $p_q$ is unique, and the interpolation is exact. For other functions, it may be shown \cite{Stoer2013} that the remainder $r_q(x) = f(x) - p_q(x)$ at any point in $[a, b]$ is given by
\begin{align}
    r_q(x) = \tilde l_q(x) \frac{f^{(q + 1)}(\xi)}{(q + 1)!}
\end{align}
for some $\xi \in [a, b]$, where $\tilde l_q(x) = \prod_{\alpha = 0}^q (x - x_\alpha)$.

The integral $\int_a^b f(x) \> \d x$ is then approximated as
\begin{align}\label{eq:app_lagrange_weights_2}
    \int_a^b f(x) \> \d x \approx \int_a^b p_q(x) \> \d x = \sum_{a = 0}^q f(x_\alpha) {\underbrace{\int_a^b l_\alpha(x) \> \d x}_{{} = w_\alpha}}.
\end{align}
Note that $\sum_{\alpha = 0}^q w_\alpha = b - a$; this can be seen by taking $f(x) = 1$, in which case the approximation is exact (because $f$ is a polynomial of degree zero). For general functions, the error is
\begin{align}
    R = \int_a^b r_q(x) \> \d x = \frac{f^{(q + 1)}(\xi)}{(q + 1)!} \int_a^b \tilde l_q(x) \> \d x.
\end{align}
Clearly, $|R|$ is bounded above by
\begin{align}\label{eq:lagrange_scalar_error_bound}
    |R| \leq \frac{(b - a)^{q + 2} \max_{\xi \in [a, b]} |f^{(q + 1)}(\xi)|}{(q + 1)!}.
\end{align}

\section{Proof of lemma \ref{lem:summation_error}}\label{app:proof_summation_error}
\begin{proof}
As before, we start from the observation that
\begin{align}
    \|(\mt U_{\eta, a}(\lambdaf, \lambdai) - \mt U_{\eta, a}^{M, q}(\lambdaf, \lambdai))\ket{n(\lambdai)}\|
    \leq \int_{\lambda_{\rm \ii}}^{\lambdaf} \d\lambda \, \|\mt A_{\eta, a}(\lambda) - \mt A_{\eta, a}^{M, q}(\lambda)\|_{n(\lambda)}.
\end{align}
By a triangle inequality, we have
\begin{align}\label{eq:interpolated_U_error_split}
    \| \mt A_{\eta, a} - \mt A_{\eta, a}^{M, q} \|_n \leq \frac12 \sum_{\kappa=1}^M \bigg( \,
    & \bigg\| \int_{\tau_{\kappa - 1}}^{\tau_\kappa} \d\tau \, \ee^{-\eta\tau} \> \partial_\lambda \H(\tau) - \sum_{\alpha = 0}^q \> w_{\kappa, \alpha} \> \ee^{-\eta \tau_{\kappa, \alpha}} \partial_{\lambda} \H(\tau_{\kappa, \alpha}) \bigg\|_n + {} \nonumber\\
    & \bigg\| \int_{\tau_{\kappa - 1}}^{\tau_\kappa} \d\tau \, \ee^{-\eta\tau} \> \partial_\lambda \H(-\tau) - \sum_{\alpha = 0}^q \> w_{\kappa, \alpha} \> \ee^{-\eta \tau_{\kappa, \alpha}} \partial_{\lambda} \H(-\tau_{\kappa, \alpha}) \bigg\|_n \,
    \bigg).
\end{align}
The first term on the right-hand side of eq. \ref{eq:interpolated_U_error_split} can be upper bounded by expanding in the eigenbasis of $\H$ and applying element-wise interpolation. For each vector element $\ee^{-\eta\tau} \bra m \dH(\tau) \ket n$, which is a scalar function of $\tau$, we can then use the error bound in eq. \ref{eq:lagrange_scalar_error_bound}. This works for our vector-valued case because $\argmax_{\sigma_m\in[\tau_{\kappa - 1}, \tau_{\kappa}]} \big| \bra m \frac{\d^{q + 1}}{\d\tau^{q + 1}} \partial_\lambda (\ee^{-\eta\tau} \H(\tau)) |_{\tau = \sigma_m} \ket n \big|$ is identical for every $m$. We obtain
\begin{align}
    \bigg\| \int_{\tau_{\kappa - 1}}^{\tau_\kappa} & \d\tau \, \ee^{-\eta\tau} \> \partial_\lambda \H(\tau) - \sum_{\alpha = 0}^q \> w_{\kappa, \alpha} \> \ee^{-\eta \tau_{\kappa, \alpha}} \partial_{\lambda} \H(\tau_{\kappa, \alpha}) \bigg\|_n \nonumber \\
    &= \bigg( \sum_m \bigg| \int_{\tau_{\kappa - 1}}^{\tau_\kappa} \d\tau \, \ee^{-\eta\tau} \> \bra m \partial_\lambda \H(\tau)\ket n - \sum_{\alpha = 0}^q \> w_{\kappa, \alpha} \> \ee^{-\eta \tau_{\kappa, \alpha}} \bra m \partial_{\lambda} \H(\tau_{\kappa, \alpha}) \ket n \bigg|^2 \bigg)^{1 / 2} \nonumber \\
    &\leq \frac{\delta\tau_\kappa^{q + 2}}{(q + 1)!} \bigg( \sum_m \max_{\sigma_m \in [\tau_{\kappa - 1}, \tau_\kappa]} \bigg| \bra m \frac{\d^{q + 1}}{\d\tau^{q + 1}} (\ee^{-\eta\tau} \partial_\lambda \H(\tau)) \Big|_{\tau = \sigma_m} \ket n \bigg|^2 \bigg)^{1 / 2} \nonumber \\
    &= \frac{\delta\tau_\kappa^{q + 2}}{(q + 1)!} \ee^{-\eta\tau_{\kappa - 1}} \bigg\| \sum_{s = 0}^{q + 1} \binom{q + 1}s (-\eta)^s [\underbrace{-\ii \H, [-\ii \H, \cdots, [-\ii \H,}_{q + 1 - s} \partial_\lambda \H] \cdots]] \bigg\|_n \nonumber \\
    &\leq \frac{\delta\tau_\kappa^{q + 2}}{(q + 1)!} \ee^{-\eta\tau_{\kappa - 1}} \sum_{s = 0}^{q + 1} \binom{q + 1}s \eta^s (2 \|\H\|)^{q + 1 - s} \|\partial_\lambda \H\|_n \nonumber \\
    &= \frac{\delta\tau_\kappa^{q + 2}}{(q + 1)!} \ee^{-\eta\tau_{\kappa - 1}} (\eta + 2 \|\H\|)^{q + 1} \|\partial_\lambda \H\|_n.
\end{align}
The calculation for the $\tau \leftrightarrow -\tau$ term in eq. \ref{eq:interpolated_U_error_split} is identical; therefore
\begin{align}\label{eq:interpolated_U_error_simpler}
    \| \mt A_{\eta, a}(\lambda) - \mt A_{\eta, a}^{M, q}(\lambda) \|_{n(\lambda)} \leq  Q^q_n(\lambda) \sum_{\kappa=1}^M \delta\tau_\kappa^{q + 2} \> \ee^{-\eta\tau_{\kappa-1}}
\end{align}
where we have defined
\begin{align}
    Q^q_n = \frac{(\eta + 2 \|\H(\lambda)\|)^{q + 1} \|\partial_\lambda \H(\lambda)\|_{n(\lambda)}}{(q + 1)!}.
\end{align}
We will now choose the $\tau_\kappa$, so as to ultimately derive the required value of $M$. Instead of using constant intervals $\delta\tau_\kappa$, it should be intuitively beneficial to pick $\delta\tau_\kappa$ small for those $\kappa$ where $\ee^{-\eta\tau}$ is large, and pick $\delta\tau_\kappa$ large when $\ee^{-\eta\tau}$ is small. An ansatz that makes every term in the sum $\sum_{\kappa=1}^M \delta\tau_\kappa^{q + 2} \> \ee^{-\eta\tau_{\kappa-1}}$ approximately constant can be found by regarding $\tau(\kappa)$ as a continuous function of $\kappa$ and solving the differential equation
\begin{align}\label{eq:tau_diff_eq}
    \frac{d\tau}{d\kappa} \ee^{-\eta\tau / (q + 2)} = \zeta
\end{align}
for some positive constant $\zeta$, subject to the boundary condition $\tau(0) = 0$. This differential equation is solved by $\tau(\kappa) = -\frac{q + 2}\eta \log(1 - \frac{\eta \kappa \zeta}{q + 2})$; the condition $\tau(M) = a$ then implies
\begin{align}
    \zeta = \frac{q + 2}{\eta M}(1 - \ee^{-\eta a / (q + 2)}) \leq \frac aM
\end{align}
so we obtain the solution
\begin{align}
    \tau(\kappa) = -\frac{q + 2}\eta \log \< \Big( 1 - \frac\kappa M (1 - \ee^{-\eta a / (q + 2)}) \Big).
\end{align}
Each term in the sum then contributes approximately $\zeta^{q + 2}$. To make this exact, we Taylor expand $\tau(\kappa)$ about $\kappa - 1$ to obtain\footnote{To be precise, we Taylor expand $\tau(x)$ about $x = \kappa - 1$ with quadratic Lagrange remainder, and this expansion is then evaluated at $\kappa$.}
\begin{align}
    \delta\tau_\kappa \> & \ee^{-\eta\tau_{\kappa - 1} / (q + 2)} - \underbrace{ \ee^{-\eta\tau_{\kappa - 1} / (q + 2)} \frac{\d\tau}{\d\kappa}(\kappa - 1) \> }_{{} = \zeta} \nonumber \\
    &= \ee^{-\eta\tau_{\kappa - 1} / (q + 2)} \frac12 \frac{\d^2\tau}{\d\kappa^2}(\hat\kappa) &\text{(for some $\hat\kappa \in [\kappa - 1, \kappa]$)} \nonumber \\
    &= \frac{q + 2}{2\eta} \ee^{-\eta\tau_{\kappa - 1} / (q + 2)} \bigg( \frac{1 - \ee^{-\eta a / (q + 2)}}{M - (1 - \ee^{-\eta a / (q + 2)}) \hat\kappa} \bigg)^2 \nonumber \\
    &\leq \frac{q + 2}{2\eta} \Big( 1 - \frac{\kappa - 1}M (1 - \ee^{-\eta a / (q + 2)}) \Big) \bigg( \frac{(1 - \ee^{-\eta a / (q + 2)})}{M - (1 - \ee^{-\eta a / (q + 2)}) \kappa} \bigg)^2 \nonumber \\
    &= \frac{q + 2}{2M\eta} \bigg( \frac{[1 - \ee^{-\eta a / (q + 2)}]^2}{[M - (1 - \ee^{-\eta a / (q + 2)}) \kappa]} + \frac{[1 - \ee^{-\eta a / (q + 2)}]^3}{[M - (1 - \ee^{-\eta a / (q + 2)}) \kappa]^2} \bigg) \nonumber \\
    &\leq \frac{(q + 2)[1 - \ee^{-\eta a / (q + 2)}]}{2M\eta} \bigg( \frac{\ee^{\eta a / (q + 2)} - 1}M + \Big( \frac{\ee^{\eta a / (q + 2)} - 1}M \Big)^2 \bigg) \nonumber \\
    &\leq \frac{(q + 2)[1 - \ee^{-\eta a / (q + 2)}]}{M\eta} = \zeta & \text{if } M \geq \ee^{\eta a / (q + 2)} - 1.
\end{align}
In the end, we find that the interpolation error obeys
\begin{align}
     \| \mt A_{\eta, a} - \mt A_{\eta, a}^{M, q} \|_n \leq M \Big( \frac{2a}M \Big)^{q + 2} Q^q_n = \frac{(2a)^{q + 2}}{M^{q + 1}} \frac{(\eta + 2 \|\H\|)^{q + 1} \|\partial_\lambda \H\|_n}{(q + 1)!}
\end{align}
implying that
\begin{align}
    \|(\mt U_{\eta, a}(\lambdaf, \lambdai) - \mt U_{\eta, a}^{M, q}(\lambdaf, \lambdai))\ket{n(\lambdai)}\|
    \leq \frac{(2a)^{q + 2}}{M^{q + 1}(q + 1)!} \int_\lambdai^\lambdaf \d\lambda \, (\eta + 2 \|\H(\lambda)\|_\infty)^{q + 1} \|\partial_\lambda \H(\lambda)\|_{n(\lambda)}.
\end{align}
We simplify the expression above by using the assumption that $\eta \leq \min_\lambda \|\H(\lambda)\|$, and by using H\"older's inequality\footnote{Define the integral norm $\|f\|_p = \big( \int_a^b \d x \, |f(x)|^p \big)^{1 / p}$ with $\|f\|_\infty = \sup_{x \in [a, b]} |f(x)|$ and let $f, g$ be such that $\|f\|_p$, $\|g\|_p$ and $\|fg\|_p$ are bounded for $1 \leq p \leq \infty$; then H\"older's inequality states that $\|fg\|_1 \leq \|f\|_p \|g\|_{p'}$ if $1 / p + 1 / p' = 1$. We choose $p = \infty$ and $p' = 1$.}:
\begin{align}
    \|(\mt U_{\eta, a}(\lambdaf, \lambdai) - \mt U_{\eta, a}^{M, q}(\lambdaf, \lambdai))\ket{n(\lambdai)}\|
    &\leq \frac{3^{q + 1}(2a)^{q + 2}}{M^{q + 1}(q + 1)!} \int_\lambdai^\lambdaf \d\lambda \, \|\H(\lambda)\|_\infty^{q + 1} \|\partial_\lambda \H(\lambda)\|_{n(\lambda)} \nonumber \\
    &\leq \frac{3^{q + 1}(2a)^{q + 2}}{M^{q + 1}(q + 1)!} \|\H\|_{\infty, \infty}^{q + 1} \|\partial_\lambda \H(\lambda)\|_{n, 1}.
\end{align}
To make this error at most $\epsilon$, it is sufficient to take
\begin{align}
    M \geq \max\bigg\{ \frac{3 \> \ee \> (2a)^{1 + 1 / (q + 1)}}{\epsilon^{1 / (q + 1)} (q + 1)} \|\H\|_{\infty, \infty} \|\partial_\lambda \H\|_{n, 1}^{1 / (q + 1)}, \ee^{\eta a / (q + 2)} - 1 \bigg\}
\end{align}
where we used Stirling's approximation to rewrite $((q + 1)!)^{-(q + 1)} \leq \ee^{1 - 1 / (q + 1)} (q + 1)^{-1} \leq \ee (q + 1)^{-1}$.
\end{proof}

\section{Proof of lemma \ref{lem:LTS_error}}\label{app:proof_LTS_error}
\begin{proof}
    We merely need to bound the sum $\sum_{\kappa=-M}^M \sum_{\alpha = 0}^q \big\| \frac{\partial^p}{\partial\lambda^p} \mt C_{\kappa, \alpha}(\lambda) \big\|$ to determine the $\tilde\Lambda$ quantity from lemma \ref{lem:Wiebe}. Clearly,
    \begin{align}
        \sum_{\kappa=-M}^M \sum_{\alpha = 0}^q \bigg\| \frac{\partial^p}{\partial \lambda^p} \mt C_\kappa(\lambda) \bigg\| &= \frac12 \|\partial_\lambda^{p+1} \H(\lambda)\| \sum_{\kappa=-M}^M \sum_{\alpha = 0}^q \delta\tau_\kappa w_{\kappa, \alpha} \, \ee^{-\eta|\tau_{\kappa, \alpha}|}.
    \end{align}
    We see that the sum on the right-hand side of this equation approximates the integral $\int_{-a}^a \d\tau \,\ee^{-\eta|\tau|}$. Considering only the positive half of the integration range, i.e. $\kappa > 0$, we have
    \begin{align}
        \sum_{\kappa = 1}^M \sum_{\alpha = 0}^q \delta\tau_\kappa w_{\kappa, \alpha} \ee^{-\eta \tau_{\kappa, \alpha}} = \sum_{\kappa = 1}^M \bigg[ \int_{\tau_{\kappa - 1}}^{\tau_\kappa} \d\tau \, \ee^{-\eta\tau} + R_\kappa \bigg] = \int_0^a \d\tau \, \ee^{-\eta\tau} + R
    \end{align}
    where $\R_\kappa \in \R$, $R = \sum_{\kappa = 1}^M R_\kappa$ and, much like in the proof of lemma \ref{lem:summation_error},
    \begin{align}
        |R_\kappa| \leq \frac{\eta^{q + 1} \> \delta\tau_\kappa^{q + 2}}{(q + 1)!} \ee^{-\eta \tau_\kappa} \leq \frac{\eta^{q + 1}}{(q + 1)!} \Big( \frac{2a}M \Big)^{q + 2}.
    \end{align}
    but since the conditions of lemma \ref{lem:summation_error} are assumed satisfied, we have
    \begin{align}
        \frac{3^{q + 1} (2a)^{q + 2}}{M^{q + 1}(q + 1)!} \|\H\|_{\infty, \infty}^{q + 1} \|\dH\|_{n, 1} \leq \epsilon \leq 1
    \end{align}
    as well as $\eta \leq \min_\lambda \|\H(\lambda)\|$, so that
    \begin{align}
        |R| \leq \frac{\eta^{q + 1}(2a)^{q + 2}}{M^{q + 1}(q + 1)!} \leq \frac{\eta^{q + 1} \epsilon}{3^{q + 1} \|\H\|_{\infty, \infty}^{q + 1} \|\dH\|_{n, 1}} \leq \frac1{3^{q + 1} \|\dH\|_{n, 1}}.
    \end{align}
    The premise that $\|\dH\|_{n, 1} \geq 3^{-(q + 1)} \frac\eta{1 - \ee^{-\eta a}}$ then directly implies that $|R| \leq \frac{1 - \ee^{-\eta a}}\eta$. The calculation for $\kappa < 0$ is identical, and we finally have
    \begin{align}
        \sum_{\kappa=-M}^M \sum_{\alpha = 0}^q \bigg\| \frac{\partial^p}{\partial \lambda^p} \mt C_\kappa(\lambda) \bigg\|  \leq \frac{2(1 - \ee^{-\eta a})}\eta \> \|\partial_\lambda^{p + 1} \H(\lambda)\|.
    \end{align}
    Therefore the condition on $\tilde\Lambda$ in eq. \ref{eq:Lambda_CD} is sufficient to fulfill the requirements of lemma \ref{lem:Wiebe}, and the result follows immediately.
\end{proof}

\section{Proof of theorem \ref{thm:gate_based_CD}}\label{app:proof_gate_based_CD}
\begin{proof}
    Consider $\tilde\U_{k, r}(\lambdaf, \lambdai)$ as in lemma \ref{lem:LTS_error}, as well as $\U_{\eta, a}(\lambdaf, \lambdai)$ for $\eta, a > 0$ as in lemma \ref{lem:regularised_truncated_unitary_error} and $\U_{\eta, a}^{M, q}$ for $M \in \N_{>0}$ as in lemma \ref{lem:summation_error}. Let $\tilde\U_{k, r}^{\rm sim}$ be the simulated version of $\tilde\U_{k, r}$, that is a version of $\tilde\U_{k, r}$ which replaces all (path-independent) operator exponentials with their (qubitisation) simulations. Clearly, if
    \begin{enumerate}[label=(\roman*)]
        \item $\|(\U_{\eta, a}(\lambdaf, \lambdai) - \U(\lambdaf, \lambdai)) \ket{n(\lambdai)}\| \leq O(\epsilon)$,
        \item $\|(\U_{\eta, a}^{M, q}(\lambdaf, \lambdai) - \U_{\eta, a}(\lambdaf, \lambdai)) \ket{n(\lambdai)}\| \leq O(\epsilon)$,
        \item $\|(\tilde\U_{k, r}(\lambdaf, \lambdai) - \U_{\eta, a}^{M, q}(\lambdaf, \lambdai)) \ket{n(\lambdai)}\| \leq O(\epsilon)$ and
        \item $\|(\tilde\U_{k, r}^{\rm sim}(\lambdaf, \lambdai) - \tilde\U_{k, r}(\lambdaf, \lambdai)) \ket{n(\lambdai)}\| \leq O(\epsilon)$
    \end{enumerate}
    then $\|(\tilde\U_{k, r}(\lambdaf, \lambdai) - \U(\lambdaf, \lambdai)) \ket{n(\lambdai)}\| \leq O(\epsilon)$ by a triangle inequality. If we set $\eta = \frac1{\sqrt2} \Delta_n^{3/2} \epsilon^{1/2} \|\dH\|_{n, 1}^{-1/2}$ and $a = \frac1\eta \log\big( 2 \> \frac{\Delta_n + \eta}{\Delta_n\epsilon\eta} \|\dH\|_{n, 1} \big)$ as in eq. \ref{eq:eta_a_conditions}, then by lemma \ref{lem:regularised_truncated_unitary_error}, condition (i) is satisfied. By assumption (c), $\min_{\lambda\in[\lambdai, \lambdaf]} \|\H(\lambda)\| \geq \eta$, so according to lemma \ref{lem:summation_error}, condition (ii) can be made true by setting parameters $\{\tau_\kappa\}$, $\{w_{\kappa, \alpha}\}$ and $M$ as in eqs. \ref{eq:tau_kappa}, \ref{eq:w_kappa,alpha} and \ref{eq:M_condition}. Furthermore, assumption (b) implies that $\|\dH\|_{n, 1} \geq 3^{-(q + 1)} \eta \geq 3^{-(q + 1)}\frac\eta{1 - \ee^{-\eta a}}$; assumption (a) implies that $\epsilon \leq (9 / 10)(5 / 3)^k \tilde\Lambda (\lambdaf - \lambdai)$; and the choice of $\gamma_p$ and $\tilde\Lambda$ ensures that eq. \ref{eq:Lambda_CD} holds. Therefore, by lemma \ref{lem:LTS_error}, condition (iii) is fulfilled if we set $r$ according to eq. \ref{eq:r_condition}. \par
    What remains is to calculate the number of quantum gates required to simulate $\tilde\U_{k, r}(\lambdaf, \lambdai)$ to precision $\epsilon$ -- that is, implement ${\tilde\U}_{k, r}^{\rm sim}(\lambdaf, \lambdai)$ such that condition (iv) is fulfilled -- given the settings of $\eta$, $a$, $M$, $q$, $k$ and $r$. For this purpose, we write $\mathcal C_\epsilon(\cdot)$ for the gate complexity of simulating a unitary to precision $\epsilon$ in spectral norm.
    We first compute the gate complexity for the simulation of the first-order, single-segment decomposition $\tilde\U_{1, 1}(\lambdaf, \lambdai)$, as given in eq. \ref{eq:tildeU1,1_CD}, and subsequently extend this to $k$\textsuperscript{th}-order, $r$-segment formulas $\tilde{\mt U}_{k, r}(\lambdaf, \lambdai)$. For this purpose, we conveniently relabel the $\kappa, \alpha$ indices with an additional subscript index $i = 1, \ldots, M(q + 1)$, as follows: $(\kappa_1, \alpha_1) = (-M, 0)$, $(\kappa_{i + 1}, \alpha_{i + 1}) = (\kappa_i, \alpha_i + 1)$ if $\alpha < q$ and $(\kappa_i + 1, 0)$ otherwise. Now, an important observation is that the $\exp[\pm \ii \mt H \tau_{\kappa, \alpha}]$ exponentials surrounding each $\exp[-\ii \mt B_{\kappa, \alpha} \delta\lambda]$ (see eqs. \ref{eq:C_B_def}--\ref{eq:exp_iC_decomp}) in the product formula partially cancel out, since
    \begin{align}
        &\exp[-\ii \mt C_{\kappa_i, \alpha_i} \delta\lambda] \exp[-\ii \mt C_{\kappa_{i + 1}, \alpha_{i + 1}} \delta\lambda] \nonumber \\
        =& \exp[-\ii \mt H \tau_{\kappa_i, \alpha_i}] \exp[-\ii \mt B_{\kappa_i, \alpha_i} \delta\lambda] \exp[\ii \mt H \tau_{\kappa_i, \alpha_i}] 
        \exp[-\ii \mt H \tau_{\kappa_{i + 1}, \alpha_{i + 1}}] \exp[-\ii \mt B_{\kappa_{i + 1}, \alpha_{i + 1}} \delta\lambda] \exp[\ii \mt H \tau_{\kappa_{i + 1}, \alpha_{i + 1}}] \nonumber \\
        =& \exp[-\ii \mt H \tau_{\kappa_i, \alpha_i}] \exp[-\ii \mt B_{\kappa_i, \alpha_i} \delta\lambda]
        \exp[-\ii \mt H (\tau_{\kappa_{i + 1}, \alpha_{i + 1}} - \tau_{\kappa_i, \alpha_i})] 
        \exp[-\ii \mt B_{\kappa_{i + 1}, \alpha_{i + 1}} \delta\lambda] \exp[\ii \mt H \tau_{\kappa_{i + 1}, \alpha_{i + 1}}].
    \end{align}
    As such, we find that the task of simulating $\tilde\U_{1, 1}(\lambdaf, \lambdai)$ to precision $\epsilon$ has gate complexity (taking $\delta\lambda = \lambdaf - \lambdai$)
    \begin{align}\label{eq:gate_complexity_U_1,1}
        \mathcal C_\epsilon(\tilde\U_{1, 1}(\lambdaf, \lambdai))
        &= 2 \bigg( \sum_{i = 1}^{M(q + 1)} \mathcal C_{\tilde\epsilon}(\exp[-\ii \mt B_{\kappa_i, \alpha_i}(\lambdai + \delta\lambda / 2) \delta\lambda / 2]) \nonumber \\
        &\quad\quad\; {} + \sum_{i = 1}^{M(q + 1) - 1} \mathcal C_{\tilde\epsilon}(\exp[-\ii \mt H(\lambdai + \delta\lambda / 2) (\tau_{\kappa_{i + 1}, \alpha_{i + 1}} - \tau_{\kappa_i, \alpha_i})] \nonumber \\
        &\quad\quad\; {} + \mathcal C_{\tilde\epsilon}(\exp[-\ii \mt H \tau_{-M, 0}]) + \mathcal C(\exp[-\ii \mt H \tau_{M, q}]) \bigg) \nonumber \\
        &\leq \sum_{\kappa = -M}^M \sum_{\alpha = 0}^q O\bigg( \frac12 \ee^{-\eta|\tau_{\kappa, \alpha}|} w_{\kappa, \alpha} \> \delta\tau_\kappa \> \|\partial_\lambda \vcg\beta(\lambdai + \delta\lambda / 2)\|_1 \> \delta\lambda + \log\frac1{\tilde\epsilon} \bigg) O(\ell) \nonumber \\
        &\quad\; {} + \sum_{i = 1}^{M(q + 1) - 1} O\bigg( \|\vcg\beta(\lambdai + \delta\lambda / 2)\|_1 (\tau_{\kappa_{i + 1}, \alpha_{i + 1}} - \tau_{\kappa_i, \alpha_i}) + \log \frac1{\tilde\epsilon} \bigg) O(\ell) \nonumber \\
        &\quad\; {} + 2 \> O\bigg(\|\vcg\beta(\lambdai + \delta\lambda / 2)\|_1 \> a + \frac1{\tilde\epsilon} \bigg) O(\ell) \nonumber \\
        &\leq O\bigg( \frac{1 - \ee^{-\eta a}}\eta \|\partial_\lambda \vcg\beta\|_{1, \infty} \> \delta\lambda + \|\vcg\beta\|_{1, \infty} a + M(q + 1) \log\frac1{\tilde\epsilon} \bigg) O(\ell)
    \end{align}
    where $\tilde\epsilon$ will be specified shortly, and we used the fact that $\frac12 \sum_{\kappa, \alpha} \ee^{-\eta|\tau_{\kappa, \alpha}|} w_{\kappa, \alpha} \delta\tau_\kappa \leq 2\frac{1 - \ee^{-\eta a}}\eta$ from the proof of lemma \ref{lem:LTS_error}. Now, to generalise this to $\tilde\U_{k, r}$, observe that a $k$\textsuperscript{th}-order LTS formula is a product of $5^{k - 1}$ first-order formulas and that an $r$-segment formula is a product of $r$ single-segment formulas. At the same time, the integration interval $\delta\lambda$ is divided into smaller subintervals such that the sum of the lengths of the subintervals over all first-order, single-segment factors is exactly $\delta\lambda$. Therefore, only the second and the third term in the last line of eq. \ref{eq:gate_complexity_U_1,1} grow with $k$ and $r$, and we obtain
    \begin{align}\label{eq:gate_complexity_U_k,r}
        \mathcal C_\epsilon(\tilde\U_{k, r}(\lambdaf, \lambdai)) 
        \leq O\bigg( \frac{1 - \ee^{-\eta a}}\eta \|\partial_\lambda \vcg\beta\|_{1, \infty} \> \delta\lambda + 5^{k - 1} r \|\vcg\beta\|_{1, \infty} a + 5^{k - 1} r M (q + 1) \log\frac1{\tilde\epsilon} \bigg) O(\ell).
    \end{align}
    We will now specify $\tilde\epsilon$. Evidently, $\tilde\U_{k, r}(\lambdaf, \lambdai)$ is a product of $O(5^{k - 1} r M (q + 1))$ (time-independent) operator exponentials. For this product to be simulated to precision $O(\epsilon)$ in operator norm, we require all constituent operator exponentials to be simulated to precision $O(\epsilon / (5^{k - 1} r M (q + 1)))$, since errors (in the sense of spectral distance) add up at most linearly by the telescoping property of the spectral norm. This means that we must set $\tilde\epsilon = \epsilon / (5^{k - 1} r M (q + 1))$ to ensure that condition (iv) is satisfied. \par
    We proceed to insert the values for $M$ from lemma \ref{lem:summation_error} (eq. \ref{eq:M_condition}) and $r$ from lemma \ref{lem:LTS_error} (eq. \ref{eq:r_condition}). Since $r$ grows superlinearly with $\tilde\Lambda$, $\tilde\Lambda$ grows with $\eta^{-1}$ and $M$ grows superlinearly with $a$, the third term in eq. \ref{eq:gate_complexity_U_k,r} is dominant in the regime of small $\eta$ and large $a$; therefore
    \begin{align}
        \mathcal C_\epsilon(\tilde\U_{k, r}(\lambdaf, \lambdai)) &\leq O\bigg( \ell 5^{k - 1} r M (q + 1) \log\frac1{\tilde\epsilon} \bigg) \nonumber \\
        &\leq \tilde O\bigg( \ell \bigg( \frac{25}3 \bigg)^k k \> (\tilde\Lambda \> \delta\lambda)^{1 + 1 / 2k} \frac{a^{1 + 1 / (q + 1)}}{\epsilon^{1 / 2k + 1 / (q + 1)}} \|\vcg\beta\|_{1, \infty} \|\partial_\lambda \H\|_{n, 1}^{1 / (q + 1)} \bigg).
    \end{align}
    Note that the $\log 1 / \tilde\epsilon$ has been absorbed into the $\tilde O(\kern0.13em\cdots)$ notation, since the gate complexity scales polynomially in $1 / \epsilon$. Lastly, we insert $a \in \tilde O(\eta^{-1})$ with the previously specified expression $\eta = \frac1{\sqrt2} \Delta_n^{3/2} \epsilon^{1/2} \|\dH\|_{n, 1}^{-1/2}$ to find
    \begin{align}\label{eq:final_CD_gate_complexity_proof}
        \mathcal C_\epsilon(\tilde\U_{k, r}(\lambdaf, \lambdai)) &\leq \tilde O\Bigg( \ell \bigg( \frac{25}3 \bigg)^k k \> (\tilde\Lambda \> \delta\lambda)^{1 + \frac1{2k}} \frac{\|\dH\|_{n, 1}^{\frac12 + \frac3{2(q + 1)}}}{\epsilon^{\frac12 + \frac1{2k} + \frac3{2(q + 1)}} \Delta_n^{\frac32 + \frac3{2(q + 1)}}} \|\vcg\beta\|_{1, \infty} \Bigg).
    \end{align}
    Finally, we observe that $\tilde\Lambda$ grows at least linearly in $\eta^{-1}$, and that $\tilde\Lambda \leq 2\eta^{-1} \Lambda$. Inserting this inequality into eq. \ref{eq:final_CD_gate_complexity_proof}, we obtain the upper bound \begin{align}\label{eq:final_CD_gate_complexity_bound_proof}
        \mathcal C_\epsilon(\tilde\U_{k, r}(\lambdaf, \lambdai)) &\leq \tilde O\Bigg( \ell \bigg( \frac{25}3 \bigg)^k k \> (\Lambda \> \delta\lambda)^{1 + \frac1{2k}} \frac{\|\dH\|_{n, 1}^{1 + \frac1{4k} + \frac3{2(q + 1)}}}{\epsilon^{1 + \frac3{4k} + \frac3{2(q + 1)}} \Delta_n^{3 + \frac3{4k} + \frac3{2(q + 1)}}} \|\vcg\beta\|_{1, \infty} \Bigg)
    \end{align}
    which is exactly eq. \ref{eq:final_CD_gate_complexity_bound}. The theorem is thus proved.
\end{proof}

\section{Sampling approach to gate-based CD \label{app:qdrift}}
In this appendix, we describe an alternative, randomised method to gate-based counterdiabatic driving. It is based on the qDRIFT algorithm \cite{Campbell2019, Berry2020}, which is a sampling protocol for time-dependent hamiltonian simulation. As in trotterisation-based CD, $\lambda$ space is discretised, in the sense that we approximate $\mt U_{\eta, a}$ as a product of unitaries
\begin{align}
    \mt U_{\eta, a}(\lambdaf, \lambdai) \approx \tilde{\mt U}_{\eta, a}(\lambdaf, \lambdai) = \prod_{j=1}^r \tilde{\mt U}_{\eta, a}(\lambda_j, \lambda_{j-1})
\end{align}
for a given partition of the interval $[\lambdai, \lambdaf]$ into $r$ subintervals $[\lambda_{j - 1}, \lambda_j]$ such that $\lambda_0 = \lambdai$ and $\lambda_r = \lambdaf$. However, the $\tilde{\mt U}(\lambda_j, \lambda_{j - 1})$ are no longer deterministically constructed, but instead sampled according to an appropriate distribution. In other words, randomised evolution is provided by a channel $\prod_{j = 1}^r \cUt(\lambda_j, \lambda_{j - 1})$. The idea of randomised gate-based CD is rooted in a Monte Carlo evaluation of the time integral expression for $\mt A_{\eta, a}(\lambda_j)$. In this setting, one would sample
\begin{align}
    \mt B_{\eta, a}(\lambda, \tau) = \ee^{-\ii \mt H(\lambda) \tau} \bigg( \frac{(1 - \ee^{-\eta a})\sgn\tau}\eta \partial_\lambda \mt H(\lambda) \bigg) \> \ee^{\ii \mt H(\lambda) \tau}, \quad P(\tau) = \begin{cases} \frac\eta{2(1 - \ee^{-\eta a})} \ee^{-\eta|\tau|} & |\tau| \leq a \\ 0 & \text{else}\end{cases}
\end{align}
so that the expected value of $\mt B_{\eta, a}(\lambda, \tau)$ over $\tau$ equals $\mt A_{\eta, a}(\lambda)$. The procedure for randomised evolution is then as follows: for each subinterval $[\lambda_{j - 1}, \lambda_j]$, we sample a $\lambda\in[\lambda_{j - 1}, \lambda_j]$ according to a distribution $p(\lambda)$ (which will be specified later) and a $\tau\in[-a, a]$ from the distribution $P(\tau)$ defined above, and apply a unitary $\tilde\U_{\eta, a}(\lambda, \tau)$ in each case. This gives rise to the channel
\begin{align}\label{eq:channel_1}
    \cUt_{\eta, a}(\lambda_j, \lambda_{j-1})(\rhom) = \int_{\lambda_{j-1}}^{\lambda_j} \d\lambda \> p(\lambda) \int_{-\infty}^{\infty} \d\tau P(\tau) \, \tilde{\mt U}_{\eta, a}(\lambda, \tau) \> \rhom \> \tilde{\mt U}_{\eta, a}\t(\lambda, \tau).
\end{align}
The intuitive choice of $\tilde\U_{\eta, a}(\lambda, \tau)$, which will be used in the algorithm, is then as follows:
\begin{align}\label{eq:channel_2}
    \tilde{\mt U}_{\eta, a}(\lambda, \tau) &= \exp[-\ii \mt B_{\eta, a}(\lambda, \tau) / p(\lambda)] \nonumber \\
    &= \exp[-\ii \mt H(\lambda) \tau] \exp \! \bigg[ \!-\!\ii \> \frac{(1 - \ee^{-\eta a})\sgn\tau}{p(\lambda)\eta} \> \partial_\lambda \mt H(\lambda)\> \bigg] \exp[\ii \mt H(\lambda)\tau].
\end{align}
We will now prove a bound on the number of quantum gates that is required to upper bound the resulting error in terms of the trace distance between the constructed channel and exact evolution. As in theorem \ref{thm:gate_based_CD}, we assume that the hamiltonian $\H$ can be decomposed as a linear combination of finitely many unitaries, $\mt H(\lambda) = \sum_{j=1}^\ell \beta_j(\lambda) \mt V_j$ (this assumption is always fulfilled in finite-dimensional Hilbert spaces).

\begin{theorem}[randomised gate-based CD]\label{thm:gate_based_CD_randomised}
    Suppose $\H(\lambda) = \sum_{i = 1}^\ell \beta_i(\lambda) \mt V_i$ is hermitian and differentiable on the interval $[\lambdai, \lambdaf]$. Define $\cUt_{\eta, a}(\lambda_j, \lambda_{j-1})$ and $\tilde{\mt U}_{\eta, a}(\lambda, \tau)$ as in eqs. \ref{eq:channel_1} and \ref{eq:channel_2} respectively, for any partition $\lambdai = \lambda_0 < \cdots < \lambda_r = \lambdaf$ of $[\lambdai, \lambdaf]$. Let $\mt U(\lambdaf, \lambdai) = \mathcal T \exp[-\ii \int_\lambdai^\lambdaf \d\lambda \, \A(\lambda)]$ and $\cU(\lambdaf, \lambdai)(\rhom) = \U(\lambdaf, \lambdai) \rhom \U\t(\lambdaf, \lambdai)$, and let $\rhom_n(\lambda) = \ketbra{n(\lambda)}$. Then there exists a partition and a distribution $p(\lambda)$ such that
    \begin{align}
        \bigg\| \, \cU(\lambdaf, \lambdai)(\rhom_n(\lambdai)) - \prod_{j=1}^r \cUt_{\eta, a}(\lambda_j, \lambda_{j-1})(\rhom_n(\lambdai)) \bigg\|_1 \leq \epsilon
    \end{align}
    and the channel $\prod_{j = 1}^r \cUt_{\eta, a}(\lambda_j, \lambda_{j - 1})$ can be implemented with
    \begin{align}
        \tilde O\bigg( \Delta_n^{-9/2} \> \epsilon^{-5/2} \> \|\partial_\lambda \mt H\|_{\infty, 1}^{7/2} \ell \> \bigg[ \max_\lambda \|\vcg\beta(\lambda)\|_1 + \|\dH\|_{\infty, 1} \max_\lambda \frac{\|\partial_\lambda \vcg\beta(\lambda)\|_1}{\|\dH(\lambda)\|} \bigg] \bigg)
    \end{align}
    quantum gates.
\end{theorem}

\begin{proof}
Define $\cU_{\eta, a}(\lambdaf, \lambdai)(\rho) = \mt U_{\eta, a}(\lambdaf, \lambdai) \rhom \mt U_{\eta, a}\t(\lambdaf, \lambdai)$ where $\U_{\eta, a}(\lambdaf, \lambdai) = \mathcal T \exp[-\ii \int_\lambdai^\lambdaf \d\lambda \> \A_{\eta, a}(\lambda)]$ and $\A_{\eta, a}(\lambda)$ is the regularised truncated AGP as in \ref{eq:regularised_truncated_AGP}. From lemma \ref{lem:regularised_truncated_unitary_error}, it immediately follows that $\|\cU(\lambdaf, \lambdai)(\rhom_n(\lambdai)) - \cU_{\eta, a}(\lambdaf, \lambdai)(\rhom_n(\lambdai))\|_1 \leq \epsilon$ if we set $\eta, a$ as in eq. \ref{eq:eta_a_conditions}. We are therefore interested in upper bounding the error
\begin{align}\label{eq:error_sampling_channel}
    \bigg\| \, \cU_{\eta, a}(\lambdaf, \lambdai)(\rhom_n(\lambdai)) - \prod_{j=1}^r \cUt_{\eta, a}(\lambda_j, \lambda_{j-1})(\rhom_n(\lambdai)) \bigg\|_1.
\end{align}
This error may be telescoped into multiple short-evolution errors:
\begin{align}
    \text{(eq. \ref{eq:error_sampling_channel})} \leq \sum_{j=1}^r \|\> \cU_{\eta, a}(\lambda_j, \lambda_{j-1})(\rhom_n(\lambda_{j-1})) - \cUt_{\eta, a}(\lambda_j, \lambda_{j-1})(\rhom_n(\lambda_{j-1}))\|_1.
\end{align}
We will now bound each term following a similar strategy to that of Berry et al. \cite{Berry2020}. First we define scaled versions of $\cU_{\eta, a}$ and $\cUt_{\eta, a}$:
\begin{align}
    \cU_{\eta, a, s}(\lambda_j, \lambda_{j-1})(\rhom) &= \mt U_{\eta, a, s}(\lambda_j, \lambda_{j-1}) \rhom \mt U_{\eta, a, s}\t(\lambda_j, \lambda_{j-1}), \nonumber \\
    \mt U_{\eta, a, s}(\lambda_j, \lambda_{j-1}) &= \mathcal P \exp \bigg[ \! -\!\ii \int_{\lambda_{j-1}}^{\lambda_j} \d\lambda \> s \> \mt A_{\eta, a}(\lambda) \bigg] ; \\[1mm]
    \cUt_{\eta, a, s}(\lambda_j, \lambda_{j-1})(\rhom) &= \int_{\lambda_{j-1}}^{\lambda_j} \d\lambda \> p(\lambda)  \int_{-\infty}^{\infty} \d\tau P(\tau) \, \tilde{\mt U}_{\eta, a, s}(\lambda, \tau) \> \rhom \> \tilde{\mt U}_{\eta, a, s}\t(\lambda, \tau), \nonumber \\
    \tilde{\mt U}_{\eta, a, s}(\lambda, \tau) &= \exp[-\ii \> s \> \mt B_\eta(\lambda, \tau) / p(\lambda)].
\end{align}
Since $\cU_{\eta, 0}(\rhom) = \cUt_{\eta, 0}(\rhom) = \rhom$, we may write (dropping the $\lambda_j, \lambda_{j-1}$ arguments for legibility)
\begin{align}
    \|\> \cU_{\eta, a}(\rhom_n) - \cUt_{\eta, a}(\rhom_n)\|_1 &= \big\|\> \big\{ \cU_{\eta, 1}(\rhom_n) - \cU_{\eta, 0}(\rhom_n) \big\} - \big\{ \cUt_{\eta, 1}(\rhom_n) - \cUt_{\eta, 0}(\rhom_n) \big\} \big\|_1 \nonumber \\
    &= \bigg\| \int_0^1 \d s \frac\d{\d s} \big( \> \cU_{\eta, a, s}(\rhom_n) - \cUt_{\eta, a, s}(\rhom_n) \big) \bigg\|.
\end{align}
For the derivatives in the last line, we use lemma 6 from Berry et al. \cite{Berry2020},
\begin{align}
    \frac\d{\d s} \mt U_{\eta, a, s}(\lambda_j, \lambda_{j-1}) = \int_{\lambda_{j-1}}^{\lambda_j} \d\lambda \, \mt U_{\eta, a, s}(\lambda_j, \lambda) [-\ii \mt A_{\eta, a}(\lambda)] \mt U_{\eta, a, s}(\lambda, \lambda_{j-1}),
\end{align}
and
\begin{align}
    \frac\d{\d s} \cUt_{\eta, a, s}(&\lambda_j, \lambda_{j-1})(\rhom) = \int_{\lambda_{j-1}}^{\lambda_j} \d\lambda \> p(\lambda) \int_{-a}^a \d\tau \, \frac\eta{2(1 - \ee^{-\eta a})} \ee^{-\eta|\tau|} \nonumber \\[1mm]
    &\times \ee^{-\ii \mt H(\lambda) \tau} \ee^{-\ii \> (1 - \ee^{-\eta a}) \> s \sgn\tau \> \partial_\lambda \mt H(\lambda) / p(\lambda) \eta} \nonumber \\
    &\times \Big[ \!-\!\<\ii \> \frac{(1 - \ee^{-\eta a})\sgn\tau}{p(\lambda) \eta} \> \partial_\lambda \mt H(\lambda), \, \ee^{\ii \mt H(\lambda) \tau} \rhom \, \ee^{-\ii \mt H(\lambda) \tau} \Big] \nonumber \\
    &\times \ee^{\ii \> (1 - \ee^{-\eta a}) \> s \sgn\tau \> \partial_\lambda \mt H(\lambda) / p(\lambda) \eta} \ee^{\ii \mt H(\lambda) \tau}.
\end{align}
We observe now that the derivatives of $\cU_{\eta, a, s}$ and $\cUt_{\eta, a, s}$ agree at $s=0$:
\begin{align}
    \frac\d{\d s} \cU_{\eta, a, s}(\lambda_j, \lambda_{j-1})(\rhom) \Big|_{s=0} = \int_{\lambda_{j-1}}^{\lambda_j} \d\lambda \, [-\ii \mt A_{\eta, a}(\lambda), \rhom] = \frac\d{\d s} \cUt_{\eta, a, s}(\lambda_j, \lambda_{j-1})(\rhom) \Big|_{s=0} \, ;
\end{align}
hence, we may invoke the fundamental theorem of calculus a second time to express
\begin{align}
    \|\> \cU_{\eta, a}(\rhom_n) - \cUt_{\eta, a}(\rhom_n)\|_1 = \bigg\| \int_0^1 \d s \int_0^s \d v \frac{\d^2}{\d v^2} \big( \> \cU_{\eta, a, v}(\rhom_n) - \cUt_{\eta, a, v}(\rhom_n) \big) \bigg\|.
\end{align}
We proceed to bound this error following Berry et al. \cite[proof of theorem 5]{Berry2020}:
\begin{align}\label{eq:sampling_error_proof}
    \|\> \cU_{\eta, a}(\rhom) - \cUt_{\eta, a}(\rhom)\|_1 &\leq \int_0^1 \d s \int_0^s \d v \, \bigg\{ \Big\| \frac{\d^2}{\d v^2} \mt U_{\eta, a, v} \> \rhom \> \mt U_{\eta, a, v}\t \Big\|_1 + 2 \Big\| \frac\d{\d v} \mt U_{\eta, a, v} \> \rhom \> \frac\d{\d v} \mt U_{\eta, a, v}\t \Big\|_1 + \Big\| \mt U_{\eta, a, v} \> \rhom \> \frac{\d^2}{\d v^2} \mt U_{\eta, a, v}\t \Big\|_1 \nonumber \\
    &\hspace{6mm} + \int_{\lambda_{j-1}}^{\lambda_j} \d\lambda \, p(\lambda) \int_{-\infty}^\infty \d\tau \> P(\tau) \nonumber \\
    &\hspace{10mm} \times \Big\| \Big[ \!-\!\<\ii \> \frac{(1 - \ee^{-\eta a})\sgn\tau}{p(\lambda)\eta} \> \partial_\lambda \mt H(\lambda), \, \Big[ \!-\!\<\ii \> \frac{(1 - \ee^{-\eta a})\sgn\tau}{p(\lambda)\eta} \> \partial_\lambda \mt H(\lambda), \, \ee^{\ii \mt H(\lambda) \tau} \rhom \, \ee^{-\ii \mt H(\lambda) \tau} \Big] \Big] \Big\|_1 \bigg\} \nonumber \\
    &\leq \int_0^1 \d s \int_0^s \d v \, \bigg\{ 4 \bigg( \int_{\lambda_{j-1}}^{\lambda_j} \d\lambda \, \|\mt A_{\eta, a}(\lambda)\|\bigg)^2 + 4 \int_{\lambda_{j-1}}^{\lambda_j} \d\lambda \, p(\lambda) \bigg( \frac{(1 - \ee^{-\eta a})\|\partial_\lambda \mt H(\lambda)\|}{p(\lambda)\eta} \bigg)^2 \bigg\}.
\end{align}
Now from the definition of $\mt A_{\eta, a}$, we find
\begin{align}
    \int_\lambdai^\lambdaf \d\lambda \, \|\mt A_{\eta, a}\| \leq \int_\lambdai^\lambdaf \d\lambda \int_{-a}^a \d\tau \> \frac12\ee^{-\eta|\tau|} \|\partial_\lambda \mt H\| = \frac{1 - \ee^{-\eta a}}\eta \|\partial_\lambda \mt H\|_{\infty, 1}
\end{align}
where $\|\partial_\lambda \mt H\|_{\infty, 1} = \int_\lambdai^\lambdaf \d\lambda \> \|\partial_\lambda \mt H(\lambda)\|$; if we now conveniently choose
\begin{align}
    p(\lambda) = \frac{\|\partial_\lambda \mt H(\lambda)\|}{\|\partial_\lambda \mt H\|_{\infty, 1}}
\end{align}
then the last line of eq. \ref{eq:sampling_error_proof} reduces to
\begin{align}
    \|\> \cU_{\eta, a}(\lambda_j, \lambda_{j-1})(\rhom) - \cUt_{\eta, a}(\lambda_j, \lambda_{j-1})(\rhom)\|_1 \leq \frac{4(1 - \ee^{-\eta a})^2}{\eta^2} \|\partial_\lambda \mt H\|_{\infty, 1}^2.
\end{align}
Lastly, we can directly follow the proof of theorem 7 from Berry et al. \cite{Berry2020} to show that
\begin{align}\label{eq:r_for_U_eta,a}
    r \geq \frac{4(1 - \ee^{-\eta a})^2}{\eta^2\epsilon} \|\partial_\lambda \mt H\|_{\infty, 1}^2
\end{align}
is sufficient to guarantee that the error from eq. \ref{eq:error_sampling_channel} satisfies
\begin{align}
    \bigg\| \, \cU_{\eta, a}(\lambdaf, \lambdai)(\rhom_n(\lambdai)) - \prod_{j=1}^r \cUt_{\eta, a}(\lambda_j, \lambda_{j-1})(\rhom_n(\lambdai)) \bigg\|_1 \leq \epsilon.
\end{align}
We thus find that one should sample and apply a unitary $\tilde\U_{\eta, a}$ $\lceil r \rceil$ times, with $r$ given by eq. \ref{eq:r_for_U_eta,a}, to achieve a total error of at most $2\epsilon$. But each of these unitaries will have to be simulated, which can be done through known hamiltonian simulation techniques.  With qubitisation \cite{Low2019b}, one can simulate an operator exponential $\exp[-\ii \mt H \theta]$ of a hamiltonian $\H = \sum_{j=1}^\ell \beta_j \mt V_j$ to error $\epsilon$ with $O([\|\vcg\beta\|_1 \> |\theta| + \log 1 / \epsilon]\ell)$ elementary gates. For the simulation error of a product of $r$ unitaries to be at most $\epsilon$, each unitary should be simulated with error at most $\epsilon / r$. Observing that $|\tau| \leq a$, $1 - \ee^{\eta a} \leq 1$, $a \in \tilde O(\eta^{-1})$ and $\eta = \frac1{\sqrt2} \Delta_n^{3/2}\epsilon^{1/2}\|\partial_\lambda \mt H\|_{\infty, 1}^{-1/2}$, we find the total gate complexity of simulating our channel to error $\epsilon$ in trace norm to be
\begin{align}
    \text{gate complexity} &\leq O\bigg( r\ell \bigg[ a \max_\lambda \|\vcg\beta(\lambda)\|_1 + \log\frac r\epsilon \bigg] \bigg) + O\bigg( r\ell \bigg[ \max_\lambda\frac{\|\partial_\lambda \vcg\beta(\lambda)\|_1}{p(\lambda) \eta} + \log \frac r\epsilon \bigg] \bigg) \nonumber \\[1mm]
    &\leq \tilde O \bigg( \frac{\|\partial_\lambda \mt H\|_{\infty, 1}^2 \ell}{\eta^3 \epsilon} \bigg[ \max_\lambda \|\vcg\beta(\lambda)\|_1 + \max_\lambda \frac{\|\partial_\lambda \vcg\beta(\lambda)\|_1 \|\dH\|_{\infty, 1}}{\|\dH(\lambda)\|} \bigg] \bigg) \nonumber\\[1mm]
    &\leq \tilde O\bigg( \Delta_n^{-9/2} \> \epsilon^{-5/2} \> \|\partial_\lambda \mt H\|_{\infty, 1}^{7/2} \ell \> \bigg[ \max_\lambda \|\vcg\beta(\lambda)\|_1 + \|\dH\|_{\infty, 1} \max_\lambda \frac{\|\partial_\lambda \vcg\beta(\lambda)\|_1}{\|\dH(\lambda)\|} \bigg] \bigg)
\end{align}
which proves the theorem.
\end{proof}


\end{document}